\documentclass[twocolumn,showpacs,pra,superscriptaddress]{revtex4}
\usepackage{dcolumn}
\usepackage{graphicx}
\usepackage{bbm}

\newcommand{\be}{\begin{equation}}
\newcommand{\ee}{\end{equation}}
\newcommand{\beq}{\begin{eqnarray}}
\newcommand{\eeq}{\end{eqnarray}}

\newcommand{\hide}[1]{}

\newcommand{\fig}[1]{Fig.\,\ref{#1}}

\newcommand{\ket}[1]{\left| #1 \right\rangle}

\begin{document}

\title{Cluster state generation using van der Waals and dipole-dipole interactions in optical lattices}

\author{Elena Kuznetsova}
\affiliation{Department of Physics, University of Connecticut,
Storrs, CT 06269}
\affiliation{ITAMP, Harvard-Smithsonian Center
for Astrophysics, Cambridge, MA 02138} 
\author{T. Bragdon}
\affiliation{Department of Physics, University of Connecticut,
Storrs, CT 06269}
\affiliation{ITAMP, Harvard-Smithsonian Center
for Astrophysics, Cambridge, MA 02138}
\author{Robin C\^ot\'e}
\affiliation{Department of Physics, University of Connecticut,
Storrs, CT 06269}
\author{S. F. Yelin}
\affiliation{Department of Physics, University of Connecticut,
Storrs, CT 06269} 
\affiliation{ITAMP, Harvard-Smithsonian Center
for Astrophysics, Cambridge, MA 02138}
\date{\today}

\begin{abstract}
We present a scalable method for generation of a cluster state for measurement-based quantum computing using van der 
Waals or dipole-dipole interactions between
neutral atoms or polar molecules in an optical lattice. Nearest neighbor entanglement is accomplished by performing a phase gate using interaction of atoms  
in Rydberg states or molecules in large dipole moment states. All nearest neighbors are sequentially entangled in a finite number of operations, independent of the 
number of qubits, producing a 1D cluster state. A universal 2D cluster state can be generated in several ms in a two-dimensional 
optical lattice by producing a series of 1D cluster states in one lattice direction, followed by application of the entangling operations in another lattice direction. 
We discuss the viability of the scheme with Rb Rydberg atoms. 

\end{abstract}

\maketitle

\section{Introduction}

Entanglement plays a major role in quantum computing \cite{QC}, quantum communication \cite{Quant-commun} and quantum metrology \cite{Quant-metrology}. 
A special type of a multipartite entangled state, 
cluster state, represents a universal resource for the measurement-based quantum computing (MBQC) paradigm \cite{1WQC}. 
In measurement-based quantum computing, quantum computations are
carried out on the cluster state through individual qubit
measurements in adaptive bases $(\ket{0}\pm e^{i\phi}\ket{1})/\sqrt{2}$.
The primary advantage of MBQC over other register-based architectures rests
in the fact that all interactions, required {\it e.g.} for two-qubit gates, are performed in
the initialization stages of the resource state. The computation may then be
performed through simultaneous measurement of many individual qubits as
warranted by the specific program being implemented.
Any one- and two-qubit gate can be realized by 
appropriate measurements, making MBQC equivalent to the standard quantum circuit model \cite{cluster-state}. Moreover, since the actual computation is done by local measurements, 
it can be faster compared to the equivalent gates in the circuit model. 

The cluster state is realized by preparing 
individual qubits in an eigenstate $\ket{+}=(\ket{0}+\ket{1})/\sqrt{2}$ of the Pauli spin operator $\sigma_{x}$, and letting the nearest neighbor qubits 
interact via (up to local operations)
an Ising Hamiltonian $H_{\rm{int}}=g(t)\sum_{\langle a,a'\rangle}\frac{1-\sigma_{z}^{(a)}}{2}\frac{1-\sigma_{z}^{(a')}}{2}$ during a time $\tau$ so that $\int_{0}^{\tau}g(t)dt=\pi$. 
It clearly shows that the cluster state can be generated by applying a phase gate $U_{\rm{PG}}=\rm{diag}(1,1,1,-1)$ between all nearest neighbor qubits.

 A cluster state was experimentally realized with photons \cite{Walther} using linear optics, but this approach is difficult to scale to a large 
number of qubits. Schemes to generate a cluster state using atom-cavity entanglement 
in the framework of cavity QED \cite{Cavity-QED}, cold ions via phonon-mediated spin-spin interactions \cite{cluster-ions}, and distributed networks of 
collectively excited atomic ensembles \cite{Ensembles}
 have also been proposed. A naturally highly scalable system is neutral atoms in an optical lattice, 
where a 1D cluster state has been produced via ultracold s-wave collisions of atoms in a spin-dependent lattice \cite{Mandel}. Neutral atoms can also be 
entangled via long-range van der Waals (vdW) or dipole-dipole interaction when 
excited to Rydberg states. Atoms in Rydberg states can have huge dipole moments of several kiloDebyes and interact strongly with each other \cite{Rydberg}, which was suggested as a 
tool to produce two-qubit gates in neutral-atom based quantum computing \cite{dip-blockade,Neut-atom-QC}. Recently, in a series of 
experiments \cite{dip-blockade-exper} Rydberg (dipole) blockade between two 
atoms has been observed, followed by the demonstrations of a blockade-assisted CNOT gate \cite{CNOT-blockade} and entanglement between two atoms \cite{blockade-entanglement}.  
It is interesting to explore the possibility to use the strong interaction in Rydberg states to generate a cluster state. Moreover, it would offer a way to produce a cluster state with 
polar molecules, coupled via dipole-dipole interaction, for which collision-based interactions can result in undesirable inelastic or chemical reaction losses.
For simplicity we primarily discuss in this work neutral atoms interacting in Rydberg states, the same scheme can be applied to molecules.

\begin{figure}
\center{
\includegraphics[width=8.5cm]{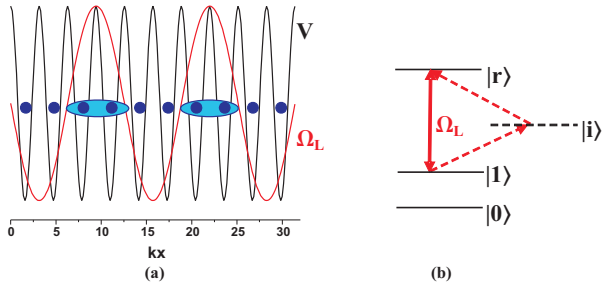}
\caption{\label{fig:superlattices} (Color online) (a) An optical lattice described by the potential $V=V_{0}\cos^{2}(kx)$, along 
with an excitation pulse in the form of a standing wave with Rabi frequency $\Omega_{L}=\Omega_{L0}\cos^{2}(kx/4+\phi)$, where $\phi=\pi/4$; (b) 
Atoms are conditionally transferred from {\it e.g.} a qubit 
state $\ket{1}$ to a Rydberg state $\ket{r}$ by optical one or 
two-photon $\pi$-pulses.}
}
\end{figure}

We propose to realize a cluster state by applying a phase gate to all pairs
of nearest neighbor atoms in an optical lattice. The phase gate can be realized 
using either direct or blockaded interaction in Rydberg states. This can be accomplished in
 a scalable way, {\it i.e.} in a finite number of operations, not 
depending on the number of atoms in the lattice. In fact, to make pair excitation controllable and to minimize errors due to 
interactions of multiple atoms in Rydberg states, it already suffices to excite every other pair: 
(a) To produce a 1D cluster state, four iterations are required and would be implemented as follows: 
using a periodic
entangler, such as a standing wave, entangle positions 1 with 2, $4n+1$ with
$4n+2$, etc.; in the second step, entangle positions 3
with 4, $4n-1$ with $4n$, etc.;  in the third step entangle positions 2 and 3, $4n-2$ and $4n-1$, etc; in the 
last step entangle positions 4 and 5, $4n$ and $4n+1$ etc. (b) For 2D cluster state generation, first the rows in
$x$-direction are entangled, then the entangling operations are applied to columns in
y-direction. This can be done using optical standing wave excitation
as is illustrated in \fig{fig:superlattices}a in the case of a 1D optical lattice. Pairs of atoms at the maxima of the standing wave intensity will get entangled, while those at the minima will 
not. Changing the phase of the standing wave fields, {\it i.e.} shifting the position of the intensity maxima and minima, will allow to perform the phase gate between all nearest neighbors, producing a cluster state.

The paper is organized as follows. In Section II we describe how a 1D cluster state can be realized using interaction in 
Rydberg states. In Section III we 
show how to generalize the technique used for the 1D cluster state generation to produce a universal 2D cluster state. Finally, we 
discuss the main features of the scheme and conclude in Section IV.  

\section{1D cluster state generation}

We start by analyzing how a 1D cluster state can be generated by applying the phase gate between all neighboring atoms in a 1D optical lattice.

\subsection{Phase gate without individual addressing of atoms in a pair (no dipole blockade)} 

\subsubsection{Gate analysis}

We assume that initially atoms are loaded 
into the ground motional state of an optical lattice described by a potential $V(x)=V_{0}\cos^{2}(kx)$, shown in \fig{fig:superlattices}a, which can be done using 
a superfluid-Mott insulator transition. 
Two ground state hyperfine sublevels $\ket{F,m_{F}}$ encode qubit states $\ket{0}$, $\ket{1}$. As a final preliminary 
step atoms are transferred into the $\ket{+}=(\ket{0}+\ket{1})/\sqrt{2}$ superposition
by applying a $\pi/2$ pulse, resonant with the qubit transition. A phase gate between atoms in neighboring sites can be implemented using strong 
vdW or dipole-dipole interaction in Rydberg states.

There are two ways to perform the phase gate depending on whether atoms are individually 
addressable or not \cite{dip-blockade}.
Standard Rydberg blockade requires individual addressing of the atoms. Individual addressing in an optical lattice has been 
demonstrated recently using a tightly focused laser beam and a spin-dependent lattice \cite{individ-address}. However, sequential application of 
the phase gate to each pair of atoms is not scalable to a large number of qubits. In a good scalable approach the cluster state has to be generated in a finite number of operations which does not 
depend on the size of the system. Below we analyze the possibility to generate the 1D cluster state assuming that atoms in each pair are not addressed 
individually.  

If one uses a standing wave excitation pulse with the Rabi frequency $\Omega_{L}=\Omega_{L0}\cos^{2}(kx/4+\phi)$, one has 
the maxima at {\it e.g.} even and minima at odd pairs sites (see \fig{fig:superlattices}a).
The atoms at the maxima are transferred from the qubit state $\ket{1}$ to the Rydberg state $\ket{r}$ either directly or by a two-photon excitation, as shown in 
\fig{fig:superlattices}b, below we assume two-photon excitation. 
The phase gate is then realized in the 
limit $\Omega_{L} \gg V_{\rm{int}}$ ($V_{\rm{int}}$ is the interaction strength in the Rydberg state $\ket{rr}$) as follows: 
1) a $\pi$-pulse resonant to the $\ket{1}-\ket{r}$ transition is applied simultaneously to both atoms, exciting each to the $\ket{r}$ state, 
since the shift of the doubly excited 
state $\ket{rr}$ is smaller than the Rabi frequency; 2) 
atoms in the $\ket{rr}$ state interact during time $T_{\rm{int}}$ and accumulate a $\pi$ phase shift $V_{\rm{int}}T_{\rm{int}}=\pi$; 3) a second $\pi$-pulse deexcites atoms back to their original 
qubit states. The standing wave therefore allows to control the Rydberg excitation pattern, which would be more difficult if a spatially homogeneous 
excitation pulse is used. We also note that vdW and dipole-dipole interactions of Rydberg atoms are long-range. As a result, when atoms are excited to $\ket{r}$, 
the interaction strength between atoms in neighboring pairs 
is comparable to the interaction strength between atoms within a pair, which will result in a phase error. The error is considerably reduced by 
exciting atoms only in every other 
pair using the standing wave. The interaction strength between atoms in closest excited even
pairs is then smaller by a factor of $\delta V_{\rm int}/V_{\rm int}=1/3^{6}\approx 10^{-3}$ for vdW and $\delta V_{\rm int}/V_{\rm int}=1/3^{3}\approx 3.7\cdot 10^{-2}$ for dipole-dipole interaction than the interaction strength within 
a pair. Here $V_{\rm int}$ is the interaction strength between atoms in a pair and $\delta V_{\rm int}$ is the interaction strength between atoms in 
closest exited pairs.

\begin{figure}
\center{
\includegraphics[width=9.cm]{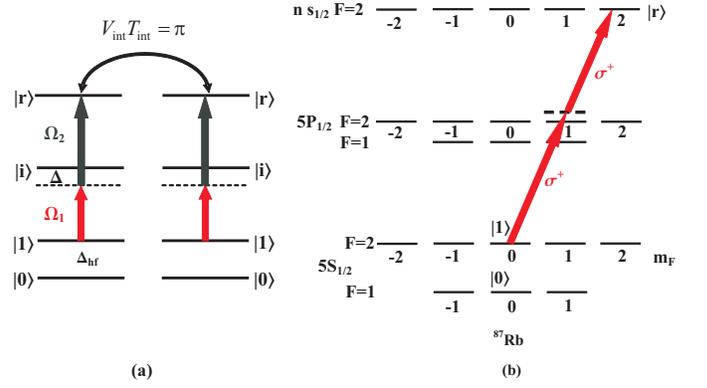}
\caption{\label{fig:level-scheme} (Color online) (a) Two-photon excitation from the $\ket{11}$ to the $\ket{rr}$ state in alkalis; (b) Level scheme of $^{87}$Rb showing qubit $\ket{0}$, 
$\ket{1}$ and Rydberg $\ket{r}$ states.}
}
\end{figure}

Excitation to Rydberg states of alkali atoms is typically a two-photon process via intermediate $p_{1/2}$, $p_{3/2}$ states. 
Using the level scheme shown in \fig{fig:level-scheme}a we can write the Schr\"odinger equation for the amplitudes of the qubit states. 
If the qubit is 
initially in the $\ket{11}$ state, the corresponding system of equations is 

\begin{eqnarray}
\label{eq:eqs_11}
i\frac{d a_{11}}{dt} & = & -\frac{2\Omega_{1}^{2}}{\Delta}a_{11}-\frac{\sqrt{2}\Omega_{1}\Omega_{2}}{\Delta}a_{+},\nonumber \\
i\frac{d a_{+}}{dt} & = & -\frac{\Omega_{1}^{2}+\Omega_{2}^{2}}{\Delta}a_{+}-\frac{\sqrt{2}\Omega_{1}\Omega_{2}}{\Delta}a_{11}-\frac{\sqrt{2}\Omega_{1}\Omega_{2}}{\Delta}a_{rr}, \nonumber \\
i\frac{d a_{rr}}{dt}&=&(V_{\rm{int}}-\frac{2\Omega_{2}^{2}}{\Delta})a_{rr}-\frac{\sqrt{2}\Omega_{1}\Omega_{2}}{\Delta}a_{+}, 
\end{eqnarray}
where $a_{11}$, $a_{+}$ and $a_{rr}$ are the amplitudes of the states $\ket{11}$, $\ket{+}=(\ket{1r}+\ket{r1})/\sqrt{2}$ and $\ket{rr}$ (other states are 
far-detuned and not populated provided that $\Delta \gg \Omega_{1},\Omega_{2}$; $\Delta$ is the detuning from the intermediate state, $\Omega_{1,2}$ are the 
Rabi frequencies of the excitation pulses). Assuming $\Omega_{1}=\Omega_{2}=\Omega$ 
and $V_{\rm{int}} \ll \Omega^{2}/\Delta$ the solution of this system is given by
\begin{eqnarray}
\label{eq:wavefunct_11}
\ket{\Psi}=e^{2i\Omega^{2}t/\Delta}\left(\frac{\ket{11}-\ket{rr}}{2}+ \right. \nonumber \\
\left. +\frac{\ket{11}-\sqrt{2}\ket{+}+\ket{rr}}{4}e^{-2i\Omega^{2}t/\Delta}+ \right. \nonumber \\
\left. +\frac{\ket{11}+\sqrt{2}\ket{+}+\ket{rr}}{4}e^{2i\Omega^{2}t/\Delta}\right), 
\end{eqnarray}
which shows that for a pulse duration $T$ such that $\Omega^{2}T/\Delta=\pi/2$ the state evolves into $\ket{\Psi}=\ket{rr}$. Next, the excitation pulses are 
switched off and the atoms interact in the doubly excited state for time $T_{\rm{int}}$ so that $V_{\rm{int}}T_{\rm{int}}=\pi$, and the state flips sign $\ket{\Psi}=-\ket{rr}$. Finally, we apply the 
same $\pi$-pulse for time $T$, bringing the system into a state $\ket{\Psi}=-\ket{11}$. 

If the initial state is $\ket{01}$ the system evolution is governed by equations (similar for the $\ket{10}$ state)

\begin{eqnarray}
\label{eq:eqs_01}
i\frac{d a_{01}}{dt}&=&-\Omega_{1}^{2}\left(\frac{1}{\Delta}+\frac{1}{\Delta+\Delta_{\rm{hf}}}\right)a_{01}-\frac{\Omega_{1}\Omega_{2}}{\Delta}a_{0r}, \nonumber \\
i\frac{d a_{0r}}{dt}&=&-\left(\frac{\Omega_{2}^{2}}{\Delta}+\frac{\Omega_{1}^{2}}{\Delta+\Delta_{\rm{hf}}}\right)a_{0r}-\frac{\Omega_{1}\Omega_{2}}{\Delta}a_{01},
\end{eqnarray}
which in the case $\Delta\gg \Delta_{\rm{hf}}$ gives the solution 
\begin{eqnarray}
\label{eq:wavefunct_01}
\ket{\Psi}&=&e^{2i\Omega^{2}t/\Delta}\left(\frac{\ket{01}-\ket{0r}}{2}e^{-i\Omega^{2}t/\Delta}+\right. \nonumber \\
&&\left. \frac{\ket{01}+\ket{0r}}{2}e^{i\Omega^{2}t/\Delta}\right), 
\end{eqnarray}
where $\Delta_{\rm{hf}}$ is the hyperfine splitting of the atomic ground state. One can see that for $\Omega^{2}T/\Delta=\pi$ the system returns to the state $\ket{\Psi}=-\ket{01}$. 

Finally, if the system is initially in the $\ket{00}$ state, the wavefunction evolves as 
\begin{equation}
\label{eq:eqs_00}
\ket{\Psi}=e^{2\Omega^{2}t/(\Delta+\Delta_{\rm{hf}})}\ket{00},
\end{equation} 
bringing the system into $\ket{\Psi}=\ket{00}$ for $\Omega^{2}T/\Delta=\pi$ if $\Delta\gg \Delta_{\rm{hf}}$. As a result, the phase gate 
$\ket{\epsilon_{1}\epsilon_{2}}\rightarrow -e^{i\pi (1-\epsilon_{1})(1-\epsilon_{2})}\ket{\epsilon_{1}\epsilon_{2}}$ is implemented.

To proceed with cluster state generation the phase gate has now to be performed with odd pairs of atoms, {\it i.e.} 
atoms 3 and 4, $4n-1$ with $4n$, etc. For that the phase $\phi$ of the excitation pulse is shifted 
by $\pi/2$ to become $\phi=3\pi/4$. At this point all atoms in even and odd pairs become entangled. As a next step the phase gate has to be applied to neighboring atoms, where 
one atom belongs to an even and another to an odd pair, {\it i.e.} to atoms 2 and 3, $4n-2$ and $4n-1$, etc. and atoms 4 and 5, $4n$ and $4n+1$, etc. which 
can be done by setting $\phi=\pi/2$ and, finally, to $\phi=\pi$. This will result in the phase gate applied to all nearest neighbors.

In \fig{fig:level-scheme}b we show the level scheme of $^{87}$Rb which we use as an example. A qubit can be encoded into 
$\ket{0}=\ket{F=1,m_{F}=0}$ and $\ket{1}=\ket{F=2,m_{F}=0}$ states, providing long coherence lifetimes due to 
small sensitivity to magnetic field fluctuations \cite{Saffman-error-analysis}. A Rydberg state $ns_{1/2}$ with 
$\ket{F=2,m_{F}=2}$ can be used as $\ket{r}$, in this case atoms will interact via isotropic vdW $V_{\rm{int}}=C_{6}/r^{6}$ interaction. 
Atoms can be excited to $\ket{r}$ using two $\sigma^{+}$ polarized pulses via intermediate $p_{1/2}$ $\ket{F=1,2,m_{F}=1}$ states. 

\subsubsection{Phase gate errors}

The fidelity of the phase gate averaged over all initial two-qubit states is calculated 
in Appendix A in detail and is given by 
\begin{eqnarray}
\label{eq:fidelity}
F= \frac{1}{4}\left[\vert \langle 00 \vert \hat{U}_{PG}\ket{00}\vert^{2}+\vert-\langle 01 \vert \hat{U}_{PG}\ket{01}\vert^{2}+\right. \nonumber \\
\left. +\vert-\langle 10 \vert \hat{U}_{PG}\ket{10}\vert^{2}+\vert-\langle 11\vert \hat{U}_{PG}\ket{11}\vert^{2}\right], 
\end{eqnarray}
where the state after the imperfect phase gate $\hat{U}_{PG}$ is compared to the state after the ideal gate $\ket{11}\rightarrow -\ket{11}$, $\ket{01}\rightarrow -\ket{01}$, 
$\ket{10}\rightarrow -\ket{10}$, $\ket{00}\rightarrow \ket{00}$. 
In Appendix A we calculate several types of intrinsic errors and find that the main errors of this type of phase gate are caused by the finite width of the ground motional 
state wavefunction of each atom in a 
lattice site. This results in a finite spread of the Rabi frequency of the standing wave excitation pulse. Other significant errors are due to  
the finite ratio of the 
interaction strength to the two-photon Rabi frequency and the decay of population in Rydberg states.  The corresponding gate fidelity is
\begin{eqnarray}
F=1-\epsilon=
1-2\pi^{2}\langle (\delta \Omega/\langle \Omega \rangle)^{2}\rangle \nonumber \\ 
-V_{\rm{int}}^{2}/(8(\Omega^{2}/\Delta)^2)-2\pi \gamma/V_{\rm{int}}-\pi \gamma/(\Omega^{2}/\Delta), \nonumber
\end{eqnarray}
where
$\epsilon$ is the corresponding gate error, and averaging is over the motional ground state wavefunction. 
In the fidelity calculation equal Rabi frequencies of the pulses $\Omega_{1}=\Omega_{2}=\Omega$ were assumed and a decay of a Rydberg state with the rate 
$\gamma$ was introduced.  

In order to find the error due to the finite width 
of the ground state motional wavefunction we 
approximate the potential at each site as harmonic with the oscillation frequency $\omega=k\sqrt{2V_{0}/m}$ and the corresponding wavefunction width 
$a=(E_{R}/V_{0})^{1/4}/k\approx 0.316/k$ for $V_{0}=100E_{R}$, which 
we use for an estimate. Here $E_{R}=\hbar^{2}k^{2}/2m$ is the atomic recoil energy, $m$ is the atomic mass. We assume the Rabi frequencies of the 
excitation pulses $\Omega_{1}=\Omega_{2}=\Omega \sim \cos(kx/4+\pi/4)$, where $x$ varies around potential minima of the n$^{\rm th}$  even pair of sites 
$x_{n}=5\pi/2(7\pi/2)+4\pi n$ 
with the Gaussian probability distribution $p(x)=\exp(-(x-x_{n})^{2}/a^2)/\sqrt{\pi}a$. 

The contribution of the first error term is 
\begin{eqnarray}
\epsilon_{\rm \Omega \;var}=2\pi^{2}\langle (\delta \Omega/\langle \Omega \rangle )^{2}\rangle = \nonumber \\
=\pi^{2}\left(ka/4\right)^{2}\tan^{2}\left(kx_{n}/4+\pi/4\right)\approx 1.05\cdot 10^{-2}. \nonumber
\end{eqnarray} 
Choosing $V_{\rm int}=3$ MHz and $\Omega^{2}/\Delta=30$ MHz, the second, third and fourth error terms are 
$\epsilon_{\rm imp\; exc}=1/8(V_{\rm{int}}/\Omega^{2}/\Delta)^2\approx 1.25\cdot 10^{-3}$, $\epsilon_{\rm Rydb \; decay1}=2\pi \gamma/V_{\rm{int}}\approx 6.7\cdot 10^{-4}$ and 
$\epsilon_{\rm Rydb \; decay2}=\pi \gamma/(\Omega^{2}/\Delta)\approx 3\cdot 10^{-5}$ for the Rydberg state lifetime $500$ $\mu$s, 
{\it i.e.} the total error is of the order of $\epsilon \approx 1.25 \cdot 10^{-2}$. The high value of the two-photon Rabi frequency $\Omega^{2}/\Delta=30$ MHz can be 
achieved using a two-photon excitation path $5s_{1/2}\rightarrow 6p_{1/2} \rightarrow ns_{1/2}$ with $422$ and $1004$ nm lasers, respectively, 
due to the larger dipole moment of the $6p_{1/2}\rightarrow ns_{1/2}$ transition \cite{Saffman-JPB}, allowing to reach a high Rabi frequency $\Omega_{2}$. 
At the same time the decay time of the $6p_{1/2}$ state (125 ns) allows to minimize the error due to the decay of the intermediate state. The probability 
of the decay during the gate is $p_{\rm se}=\pi \gamma_{6p_{1/2}}/\Delta$. Choosing $\Delta=40$ GHz the probability of the intermediate state 
decay is $p_{\rm se}\sim 10^{-4}$.

There is also an error due to the undesirable interaction of atoms belonging to different excited pairs, $\epsilon_{\rm dif \; pairs}=(3\pi^{2}/16)\left(1/8+19\pi^{2}/256\right)(\delta V_{\rm int}/V_{\rm int})(V_{\rm int}/(\Omega^{2}/\Delta))^{3}$ 
(see Appendix A, section A). Assuming that interacting atoms belong to the closest excited pairs, we have $\delta V_{\rm int}/V_{\rm int}=1/3^{6} \approx 10^{-3}$ 
for vdW and $\delta V_{\rm int}/V_{\rm int}=1/3^{3}\approx 3.7\cdot 10^{-2}$ for dipole-dipole interaction. As a result, the error is 
$\epsilon_{\rm dif \; pairs}\approx 1.6\cdot 10^{-6}$ for vdW and $\epsilon_{\rm dif \; pairs}\approx 5.9\cdot 10^{-5}$ 
for dipole-dipole interaction.

This analysis assumes that the excitation from $\ket{1}$ to $\ket{r}$ and back is adiabatic with respect to the lattice motional frequency and 
there is no uncertainty in the interaction strength, analyzed in Appendix A, section C. If the excitation-deexcitation is non-adiabatic, there is an 
additional gate error due to the uncertainty in the interaction strength $\epsilon_{\rm non \;adiab}=(\pi^{2}/4)(a/R)^{2}-(\pi^{2}/4)(a/R)^{4}(V_{\rm int}/\omega)^{2}$, where $R$ is the 
distance between nearest neighbors in the lattice. This error was derived assuming $V_{\rm int}\le \omega$, which is not the case in our estimates. If 
$V_{\rm int} > \omega$ the error will be higher, and we use the first term $\epsilon_{\rm non \; adiab}=(\pi^{2}/4)(a/R)^{2}\approx 2.5\cdot 10^{-2}$ as a lower bound for this type of error. 
The error derivation assumes that the trapping potential for the ground and Rydberg 
state is the same, providing the same motional frequency. This can be achieved in a 
blue-detuned optical lattice at a "magic" wavelength, at which ground and Rydberg state polarizabilities are 
equal \cite{Saffman-magic-wavelength}. On the other hand, the error can be avoided if the trapping 
lattice is switched off when atoms are excited to Rydberg states and the gate pulses are much shorter 
than the motional period. In this case atoms move only a small fraction of the ground motional state 
width during the gate and will be recaptuted in the ground motional state once the trapping lattice 
is switched back on.

Finally, there is a finite probability to excite atoms in "inactive" lattice sites, sitting in minima of the standing wave excitation pulse. The probability 
that the pair of atoms stays in the initial state after the gate, averaged over four possible initial states, is calculated in Appendix A, section E. For 
a two-photon excitation the averaged probability $\langle P \rangle \approx 0.75$.

The error analysis shows that first, vdW interaction results in smaller errors caused by the interaction of atoms in 
different excited pairs compared to the dipole-dipole one. Second, the excitation to Rydberg states has to be adiabatic 
to avoid errors due to the uncertainty of the interaction strength. On the other hand, these errors could be avoided altogether if excitation to 
the $\ket{rr}$ state is not required. Next we therefore analyze the phase gate based on Rydberg (dipole) blockade, where only one atom is excited to a Rydberg state, 
provided individual addressing of atoms in a pair is possible. 

\subsection{Phase gate with "individual" addressing: dipole blockade} 

\subsubsection{Gate analysis}

 If individual addressing of atoms in a pair is possible, one can use standard dipole blockade \cite{dip-blockade}, 
and the phase gate is realized in the limit $\Omega_{L} \ll V_{\rm{int}}$ as follows: 
1) a control atom is excited 
from one of the qubit states, {\it e.g.} a qubit state $\ket{1}$ to a Rydberg state $\ket{r}$ by a $\pi$-pulse; 2) a $2\pi$-pulse of the same frequency 
is applied to a target atom. Dipole-dipole or vdW 
interaction shifts the energy of the doubly excited state $\ket{rr}$, as a result 
the $2\pi$-pulse has no effect on the target atom (the excitation is blockaded) since the 
Rabi frequency of the pulse is much smaller than the energy shift; 3) the control atom is deexcited back to its original qubit state. We stress that 
we do not require individual addressing of atoms in the lattice, we only need atoms in each pair to be separately addressable, this is why we call 
it "individual" addressing.

Individual addressing in a pair can be realized using a polarization gradient optical lattice \cite{polariz-gradient}, where 
not only the intensity of the lattice field but also its polarization changes periodically in space.  
A lattice of double-wells was demonstrated in \cite{Double-well-latt} with the polarization being linear in one site and 
elliptical in the neighboring site of each well. The elliptical polarization 
results in a non-zero contribution from a vector part of polarizability, producing a state-dependent shift $\sim \alpha_{v}(\vec{E}_{L}^{*}\times \vec{E}_{L})\vec{F}$, 
where $\alpha_{v}$ is the vector part of the polarizability, $\vec{F}$ is the atomic angular momentum, and $\vec{E}_{L}$ is the positive frequency part 
of the total electric field $\vec{E}=\vec{E}_{L}e^{-i\omega t}+c.c.$. 
The contribution from the vector part can, therefore, be viewed as a fictitious magnetic field 
$\vec{B}_{\rm{fict}}\sim \alpha_{v}(\vec{E}_{L}^{*}\times \vec{E}_{L})$.

\begin{figure}
\center{
\includegraphics[width=9.cm]{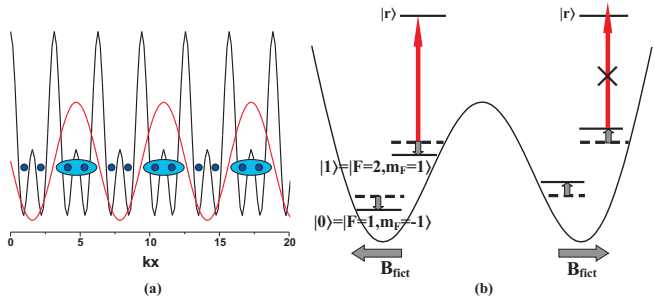}
\caption{\label{fig:double-well} (Color online) (a) Double-well lattice described by the potential $V=V_{0}\cos^{2}kx+V_{1}\cos^{2}2kx$ ($V_{1}=V_{0}$) along 
with a standing wave excitation pulse with Rabi frequency $\Omega_{L}=\Omega_{L0}\cos^{2}(kx/2+\pi/4)$; (b) Shift of qubit states 
$\ket{0}=\ket{F=1,m_{F}=-1}$ and $\ket{1}=\ket{F=2,m_{F}=1}$ in left and right sites of a double-well in a polarization gradient lattice described in the 
text.}
}
\end{figure}

Atoms can be loaded into a double-well optical lattice, formed by 
two standing wave fields of different polarizations. Let us consider the lattice field 
\begin{eqnarray}
\vec{E}_{L}=E_{0}\vec{e}_{y}e^{ikx+i\phi}+E_{0}\vec{e}_{y}e^{-ikx-i\phi}+ \nonumber \\
+iE_{1}\vec{e}_{z}e^{2ikx}+iE_{1}\vec{e}_{z}e^{-2ikx} \nonumber.
\end{eqnarray}
This lattice can be produced by two pairs of laser beams, intersecting at $\pi$ and $\pi/3$ angles \cite{Tommaso}. 
The corresponding lattice potential is a sum of scalar and vector parts $V=-\alpha_{s}|\vec{E}_{L}|^{2}/4+i\alpha_{v}(\vec{E}_{L}^{*}\times \vec{E}_{L})\vec{F}/4=V_{s}+V_{v}$, 
where $\alpha_{s}$ is the scalar
polarizability. The scalar part $V_{s}=V_{0}\cos^{2}(kx+\phi)+V_{1}\cos^{2}2kx$ (where $V_{0}=-\alpha_{s}E_{0}^{2}$ and 
$V_{1}=-\alpha_{s}E_{1}^{2}$), represents a 
double-well lattice, with a spatial period $a=\pi/k$; the minima of the n$^{\rm th}$ double-well are at 
$(kx_{\rm{min}})_{n}=(\arcsin(V_{0}/4V_{1})/2+\pi/4+\pi n, -\arcsin(V_{0}/4V_{1})/2+\pi/4+\pi (2n+1)/2)$.
The vector part is given  by 
\begin{eqnarray}
V_{v}=-2\alpha_{v}\sqrt{V_{0}V_{1}}\cos(2kx) {\rm cos}(kx) F_{x}/\alpha_{s}, 
\end{eqnarray} 
and at the minima of a double-well 
$V_{v}=\pm \alpha_{v}\sqrt{V_{0}V_{1}(1-V_{0}/4V_{1})/2}V_{0}F_{x}/2V_{1}\alpha_{s}$. Here $F_{x}$ is the projection of the 
total angular momentum $\vec{F}$ on the direction of the fictitious magnetic field $\vec{B}_{\rm fict}$ (x-direction), {\it i.e.} $F_{x}=m_{F}$; the 
$\pm$ signs refer to the left and right sites of the well. 
The scalar part of the potential is typically larger than the vector part, since the vector 
polarizability is about an order of magnitude smaller than the scalar part for alkalis.
Assuming $\alpha_{v}/\alpha_{s}\sim 0.1$, $V_{0}=V_{1}$ and $V_{0}=100E_{R}$ the vector shift at the well's minima is $V_{v}\approx \pm 3E_{R}F_{x}$. 
We can encode a qubit into the states $\ket{0}=\ket{F=1,m_{F}=-1}$, $\ket{1}=\ket{F=2,m_{F}=1}$, this qubit transition is insensitive 
to magnetic field fluctuations at a bias magnetic field $B_{\rm bias}=3.23$ G \cite{Saffman-error-analysis}. As a result, if a magnetic field is applied, 
the $\ket{0}$ and $\ket{1}$ states will shift by the same amount preserving the qubit frequency. 
The $V_{v}$ term, acting as a fictitious magnetic field, will then
shift states $\ket{0}$, $\ket{1}$ in the right site with respect to the qubit states in the left site of the well \cite{Individ-address-double-well} (see \fig{fig:double-well}b),   
 allowing to selectively excite only one atom in each pair, while having another atom unaffected by the excitation pulse. 
For $^{87}$Rb $E_{R}=3.5$ kHz 
at the lattice wavelength $\lambda=810$ nm, and the shifts for the $\ket{0}$, $\ket{1}$ states are $V_{v}\sim \pm 10$ kHz. Larger shifts of the order of 
$50-100$ kHz \cite{Individ-address-double-well} can be realized 
with optimized parameters of the lattice.

If the two-photon excitation Rabi frequency is much smaller than the vector 
shift $\Delta_{\rm{vec}}=2|V_{v}|$, the atoms in the left and 
right sites can be excited to the $\ket{r}$ state selectively.
The phase gate then can be realized using dipole blockade as: 1) a $\pi$-pulse resonant to the $\ket{1}-\ket{r}$ transition of a 
control atom is applied, exciting it to the $\ket{r}$ state; a target atom is off-resonant by $\Delta_{\rm{vec}}$, which is the relative shift 
of the qubit state $\ket{1}$ in left and right site, and is not excited; 2) a $2\pi$-pulse resonant to 
the $\ket{1}-\ket{r}$ transition of the target atom is applied. The doubly excited $\ket{rr}$ state is shifted by the large interaction energy $V_{\rm{int}}$ and 
is not populated. The control atom is off-resonant by $\Delta_{\rm{vec}}$ and is not affected; 3) finally, a $\pi$-pulse resonant to the control atom 
brings it back to its original qubit state. These steps are analyzed in detail in Appendix B and are summarized in the table below, which shows the evolution of the two-qubit states:

\begin{equation}
\label{eq:dipole-blockade}
\begin{array}{r@{\quad\stackrel{\pi_c}{\rightarrow}\quad}r@{\quad\stackrel{2\pi_t}{\rightarrow}\quad}r@{\quad\stackrel{\pi_c}{\rightarrow}\quad}r}
\ket{00} & -\ket{00} & -\ket{00} & \ket{00}, \\
\ket{01} & -\ket{01} & \ket{01} & -\ket{01}, \\
\ket{10} & -i\ket{r0} & -ie^{i\theta}\ket{r0} & -e^{i\theta}\ket{10}, \\
\ket{11} & -i\ket{r1} & -ie^{i\theta}\ket{r1} & -e^{i\theta}\ket{11} \nonumber,
\end{array}
\end{equation}
where $\theta=\pi \Delta_{\rm{vec}}/\Omega^{2}/\Delta$, the first qubit corresponds to the control and the second to the target atom. One can see that 
the phase gate can be realized provided that $\theta=2\pi n$.

\subsubsection{Phase gate errors}

The averaged fidelity of the phase gate with dipole blockade was calculated in Appendix B and is given by 
\begin{eqnarray}
F=1-\epsilon=1-2\pi^{2}\langle (\delta \Omega/\langle \Omega \rangle)^{2}\rangle -7\pi \gamma/(4\Omega^{2}/\Delta) \nonumber \\
-(\Omega^{2}/\Delta)^{2}/(2\Delta_{\rm{vec}}^{2}). \nonumber
\end{eqnarray} 
The error terms are 
due to the spread of the two-photon Rabi frequency experienced by each atom 
\begin{eqnarray} 
\epsilon_{\rm \Omega \; var}=2\pi^{2}\langle (\delta \Omega/\Omega)^{2}\rangle= \nonumber \\
=\pi^{2}(ka)^{2}\frac{1-\sqrt{(1+V_{0}/4V_{1})/2}}{1+\sqrt{(1+V_{0}/4V_{1})/2}} \approx 0.15, \nonumber
\end{eqnarray} 
decay of Rydberg states $\epsilon_{\rm Rydb \; decay}=7\pi \gamma/(4\Omega^{2}/\Delta) \approx 4.38 \cdot 10^{-2}$, and the imperfect frequency 
selectivity between the control and target 
atoms 
$\epsilon_{\rm imp \; block}=(\Omega^{2}/\Delta)^{2}/(2|\Delta_{\rm{vec}}|^{2}) \approx 0.02$. 
Here we assumed $V_{\rm int} \gg \Delta_{\rm vec} \gg \Omega^{2}/\Delta$, $V_{0}=V_{1}$, $a=(E_{R}/V_{0})^{1/4}/k\approx 0.316/k$ for $V_{0}=100E_{R}$, $1/\gamma=500$ $\mu$s, the two-photon Rabi frequency $\Omega^{2}/\Delta=40$ kHz, and the vector shift of the qubit states 
$\Delta_{\rm{vec}}=200$ kHz. It shows that the total error of the phase gate in this case is $\epsilon \approx 0.21$ and the phase gate time is 
$T_{PG}=2\pi/\Omega^{2}/\Delta\approx 25$ $\mu$s. 

Let us also discuss the error due to undesirable excitation of atoms in "inactive" wells, where the Rabi frequency is close to a minimum. 
It was calculated in detail in Appendix B, section D, and the probability to find the pair of atoms in the initial state after the gate, averaged 
over all four initial states, is $\langle P \rangle \approx 0.998$ in the case of two-photon excitation to Rydberg states.

By changing the phase of 
the standing wave so that intensity maxima shift to odd double-wells the phase gate can be realized in odd pairs. At this stage all pairs in double-wells 
are entangled. In the next 
subsection we describe how the phase gate can be realized with neighboring atoms belonging to different double-wells.

\subsubsection{Lattice manipulation} 

\begin{figure}
\center{
\includegraphics[width=9.5cm]{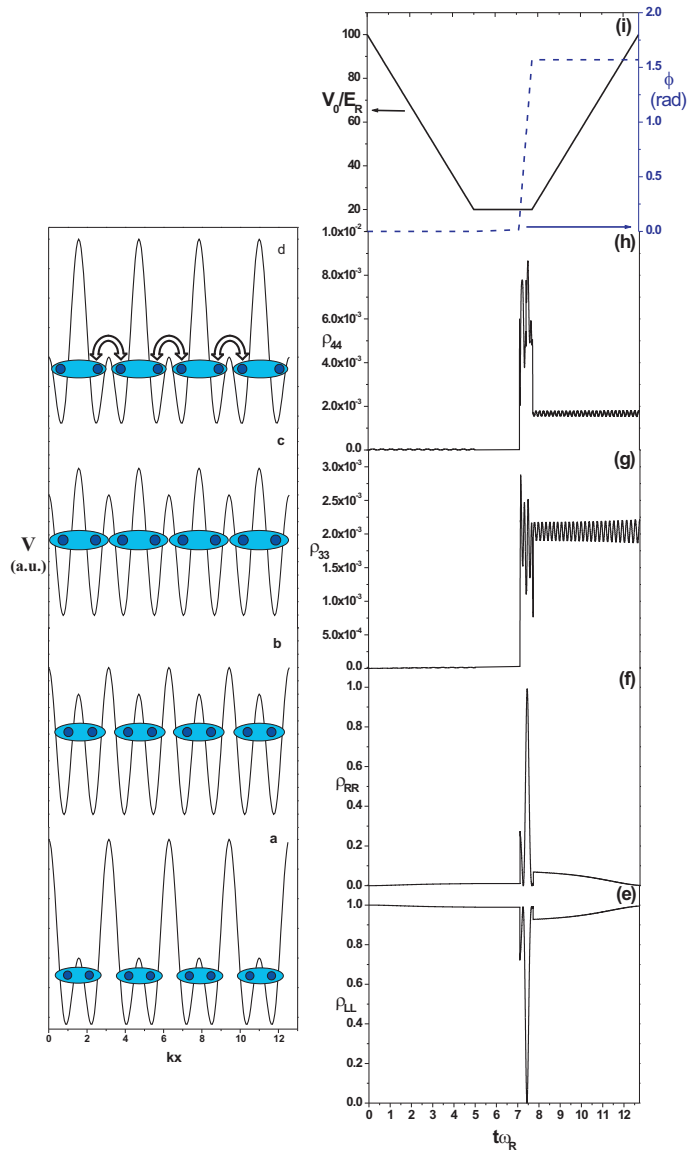}
\caption{\label{fig:Superlatt-manip} (Color online) Left panel: (a) Double-well lattice described by the potential $V=V_{0}\cos^{2}(kx+\phi)+V_{1}\cos^{2}(2kx)$, $V_{1}=V_{0}$, $\phi=0$; 
(b) $V_{0}$ is ramped down to $V_{0}=0.2V_{1}$; (c) The lattice phase is shifted from $0$ to $\phi=\pi/2$, $V_{0}=0.2V_{0}$; 
(d) $V_{0}$ is ramped up to $V_{0}=V_{1}$; Right panel: populations of the four lowest motional states during the $V_{0}$ and $\phi$ manipulations.}
}
\end{figure}

In the previous subsection we showed how a phase gate can be applied to pairs of atoms in double-wells. To proceed with the cluster-state generation 
the phase gate has to be realized with the neighboring atoms in different double-wells, {\it i.e.} between 
each atom in a right site of the $n^{\rm{th}}$ double-well and an atom in a left site of the $(n+1)^{\rm th}$ double-well. 
The phase gate operations described in the previous subsection can be applied if the atoms are brought to the same double-well. This can be 
achieved by adiabatically manipulating the lattice in the following way: (i) decreasing $V_{0}$ (\fig{fig:Superlatt-manip}b, 
left panel) which raises the barrier in each double-well, (ii) ramping the phase $\phi$ from $0$ to $\pi/2$ 
(\fig{fig:Superlatt-manip}c, left panel), (iii) finally, increasing $V_{0}$ back to its initial value. As a result, the atoms 
that were in the right and left sites of neighboring double-wells end up in the 
same double-well (\fig{fig:Superlatt-manip}d, left panel). 

\begin{figure}
\center{
\includegraphics[width=9.5cm]{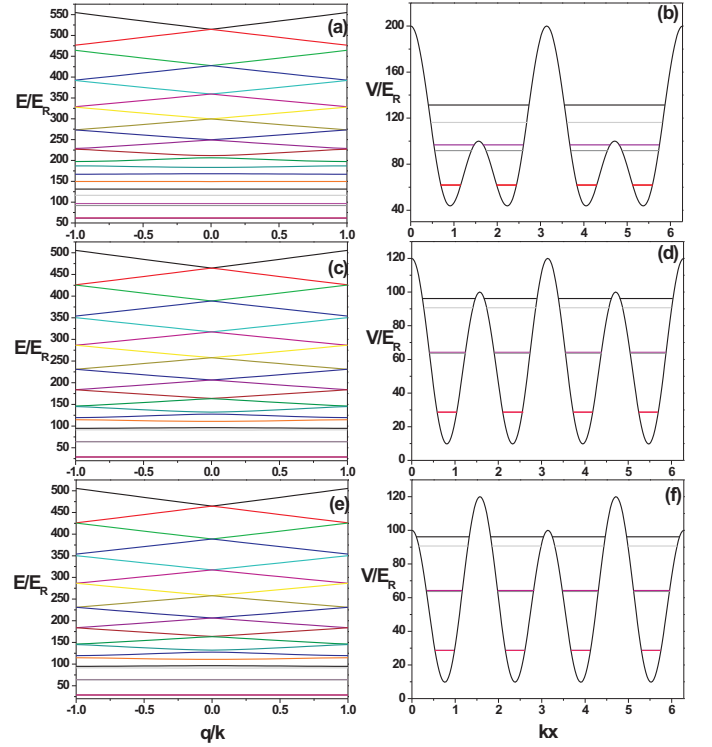}
\caption{\label{fig:Bloch-bands} (Color online) Left panel: Bloch bands of the double-well lattice $V=V_{0}\cos^{2}(kx+\phi)+V_{1}\cos^{2}(2kx)$, (a) $V_{1}=V_{0}=100E_{R}$, 
$\phi=0$; (c) $V_{0}=0.2V_{1}$, $\phi=0$; (e) $V_{0}=0.2V_{0}$, $\phi=\pi/2$; Right panel: Double-well potential along with lowest Bloch energies, where 
$V_{0}$ and $\phi$ in (b), (d) and (f) are the same as in the left panel for (a), (c) and (e).}
}
\end{figure}

The lattice manipulation has to be adiabatic to avoid motional excitation of atoms.
Modeling the evolution of the atomic motional state during the lattice manipulation requires calculation of Bloch bands and Bloch functions for every configuration of the lattice. 
These are the eigenstates and eigenfunctions of the single-particle atomic Hamiltonian,
which can be found by solving the Schr\"odinger equation with the potential $V(x,t)$.
The potential $V(x,t)$ is periodic, and can be written as a discrete Fourier sum, containing terms with wavevectors $k_{x}=\pm k,\pm 2k$, in the 
following way:
\begin{eqnarray}
V(x,t) & = & (V_{0}+V_{1})/2+V_{0}(e^{2ikx+2i\phi}+e^{-2ikx-2i\phi})/2+ \nonumber \\
&& +V_{1}(e^{4ikx}+e^{-4ikx})/2 \nonumber,
\end{eqnarray}
allowing one to write the solutions of the Schr\"odinger equation in the form of Bloch functions $\psi_{q}^{(n)}(x)=e^{iqx}u_{q}(x)$.  Here $q$ is the quasi-momentum 
(restricted to the first Brillouin zone), $n$ is the band index, and  
\begin{equation}
\label{eq:Bloch-func}
u_{q}(x)=\sum_{n=-N_{\rm{max}}}^{N_{\rm{max}}}c_{n}(q)e^{2inkx},
\end{equation}
with $N_{\rm{max}}$ a suitable cutoff number.
The resulting system of equations for the $c_{n}$ coefficients and eigenenergies $E^{(n)}(q)$  
\begin{eqnarray}
\frac{\hbar^{2}(q+2nk)^{2}}{2m}c_{n}+\frac{V_{0}}{2}(c_{n+1}e^{-2i\phi}+c_{n-1}e^{2i\phi})+ \nonumber \\
+ \frac{V_{1}}{2}(c_{n+2}+c_{n-2})=(E^{(n)}(q)-\frac{V_{0}+V_{1}}{2})c_{n} \nonumber,
\end{eqnarray}
is solved numerically by truncating the sum in Eq.(\ref{eq:Bloch-func}) at some $N_{\rm{max}}$ providing a necessary precision for the eigenenergies. We 
used $N_{\rm{max}}=10$ to calculate the lowest Bloch energies and functions. Different configurations of the double-well 
potential and the corresponding band structures are shown in \fig{fig:Bloch-bands}. The left panel demonstrates Bloch bands corresponding to the three 
stages of lattice manipulation, while in the right panel of the 
figure the energies of the $q=0$ eigenstates of several lowest bands are shown. The energies of the lowest bands weakly depend on $q$, as can be 
seen from \fig{fig:Bloch-bands}(a),(c),(e), and the energies and Bloch wavefunctions corresponding to $q=0$ can be used.
Given the Bloch functions $\psi_{q}^{(n)}$, which are delocalized over the entire lattice, one can construct Wannier functions which are localized at lattice sites $x_{i}$
\begin{equation}
w^{(n)}(x-x_{i})=\frac{1}{\sqrt{N}}\sum_{q}e^{-iqx_{i}}\psi_{q}^{(n)}(x) \nonumber.
\end{equation}

\begin{figure}
\center{
\includegraphics[width=9.5cm]{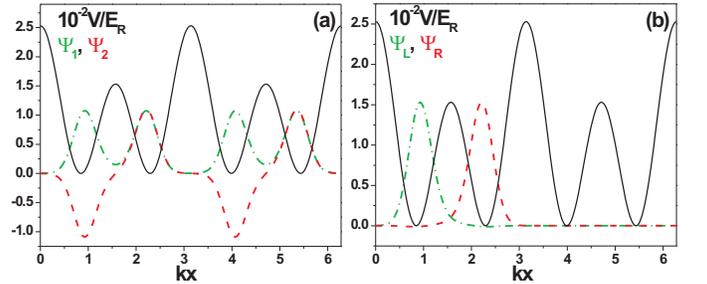}
\caption{\label{fig:Wannier-functions} (Color online) (a) Bloch functions $\psi_{1}$ (green dot-dashed curve) and $\psi_{2}$ (red dashed curve) of the first and second Bloch bands along 
with the lattice potential $10^{-2}V$; (b) Wannier functions centered in the left $\psi_{L,i}=(\psi_{1,i}-\psi_{2,i})/\sqrt{2}$ and right 
$\psi_{R,i}=(\psi_{1,i}+\psi_{2,i})/\sqrt{2}$ sites of a double-well. }
}
\end{figure}

In a double-well lattice the two lowest bands are separated in energy by much less ($\le 0.1E_{R}$) than the typical motional excitation energy $\sim \sqrt{4V_{0}E_{R}}$. 
As a result, even at ultracold temperatures both bands are going to be populated. In this situation to obtain a correct description of the system 
evolution generalized Wannier functions are introduced \cite{GWF}, which are superpositions of the Wannier functions of different energy bands. In our 
case the Bloch functions for the first two bands are symmetric (ground band) and anti-symmetric (second band) around the center of a double-well (shown 
in \fig{fig:Wannier-functions}a). Combining the Wannier functions corresponding to the two bands as $\psi_{L,i}=(\psi_{1,i}-\psi_{2,i})/\sqrt{2}$ and 
$\psi_{R,i}=(\psi_{1,i}+\psi_{2,i})/\sqrt{2}$, one can obtain generalized Wannier functions 
localized in the left and right well, respectively (see \fig{fig:Wannier-functions}b).

The lattice parameters have to be changed adiabatically to avoid undesirable motional excitations of the atoms, 
{\it i.e.}, slow compared to the lattice motional energies.
We checked the adiabaticity of lattice manipulation by numerical modeling of the system evolution. 
We took into account first four Bloch states to simplify the analysis and  assumed that initially atoms are in the ground 
Bloch state. As we already mentioned, the lowest Bloch bands weakly depend on the quasi-momentum $q$, and, as a result, one can approximate the corresponding Wannier 
functions by the Bloch functions $\psi_{L,i}$ and $\psi_{R,i}$, restricted to a single site. The population evolution is shown in the right panel of 
\fig{fig:Superlatt-manip}, along with the time-dependence of 
the amplitude $V_{0}$ and phase $\phi$.
 By optimizing the manipulation time we found that $99.92\%$ of the population stays in the ground Bloch band. We note that this value can be further increased 
by using more complex 
(optimized) functions $V_{0}(t)$ and $\phi(t)$. The total time required for the manipulation is $\approx 600$ $\mu$s.

We can now estimate the 
time required to generate a 1D cluster state. In the case considered in Section II A, when atoms are not addressed individually, the phase gate has to be 
performed adiabatically with the gate time $T_{\rm{PG}}\gg 1/\omega \approx 1$ $\mu$s, as a result, the excitation-deexcitation pulses have to be longer than $10$ $\mu$s. 
We can assume the phase gate duration $T_{\rm{PG}}\approx 20$ $\mu$s. The resulting time of cluster state 
generation, including four phase gate sequences is then $T_{1D}\approx 80$ $\mu$s. If atoms in a pair are individually addressable, the gate 
duration was found to be $\approx 25$ $\mu$s in Section II B. The total  time including the lattice manipulation is then $T_{1D}\approx 700$ $\mu$s.

\section{2D cluster state generation}

The scheme described in the previous section can be extended to generate a 2D cluster state, which is required for universal quantum computation. 
First, atoms can be loaded in a 2D lattice with 
$V=V_{0}\cos^{2}(k_{1}x+\phi)+V_{1}\cos^{2}(2k_{1}x)+V_{2}\cos^{2}(2k_{2}y)$, which produces a regular lattice with a period $\pi/k_{2}$ in $y$ 
direction and a regular or double-well lattice in $x$ direction. The $y$-lattice period is assumed sufficiently large 
so that atoms in neighboring $x$-chains 
do not interact when excited to Rydberg states. In this way, following the steps of Sections II A or II B, we can produce a series of 1D cluster states in the $x$ direction. 
As a next step we adiabatically reduce $V_{1}$ to get a regular lattice in the $x$-direction with a period $\pi/2k_{1}$ (in the case of a double-well lattice), 
followed by stretching of 
the $x$-lattice. Next, the $y$-lattice period is reduced to bring the $x$-chains closer and the regular or double-well lattice 
$V=V_{2}\cos^{2}(2k_{2}y+\phi)+V_{3}\cos^{2}(k_{2}y+\phi)$ is switched on in the $y$ direction. Next, 
the entanglement operations of Sections II A or II B can be repeated for 1D chains in the $y$ direction, producing a 2D cluster state.

Let us estimate the time required to generate the 2D state. As we showed in the end of Section II C a 1D cluster state in the $x$-direction requires 
$T_{\rm{1D}}\approx 80$ $\mu$s without and $T_{\rm{1D}}\approx 700$ $\mu$s with individual addressing in a pair to be realized. 
Adiabatic reduction of $V_{0}$ will take $\approx 250$ $\mu$s, as can be seen from \fig{fig:Superlatt-manip}(e) in the right panel. Next, we need to estimate 
the time required to stretch the $x$-lattice. The stretch has to be adiabatic for atoms to remain in the ground motional state of the lattice. We 
modeled the stretch of the lattice $V=V_{1}\cos^{2}(kx)$ so that the $k$ vector was adiabatically changed from $2k_{1}$ to $0.4k_{1}$. The 
period of the lattice $5\pi/2k_{1}$ at the end of the stretch is comparable to the distance $3\pi/2k_{1}$ between atoms which belong 
to every other excited pair in the double-well. As a result, the error due to the interaction of atoms in neighboring $x$-chains is smaller than the error due to the interaction 
of atoms in neighboring excited pairs in the same chain.

\begin{figure}
\center{
\includegraphics[width=6.5cm]{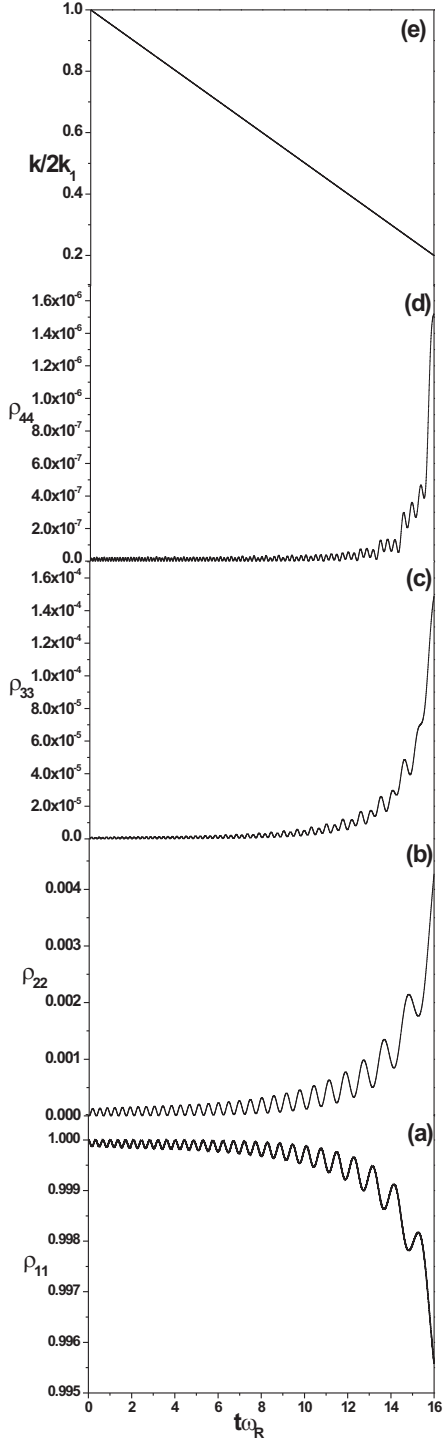}
\caption{\label{fig:Latt-stretch} (a)-(d) Populations of the four lowest Bloch bands during the lattice stretch; (e) the $k$ vector is linearly reduced 
from $2k_{1}$ to $0.4k_{1}$.}
}
\end{figure}

We modeled the evolution of the system by calculating the Bloch bands and Bloch functions, and used the latter to construct single-site Wannier functions of the lattice $V(x,t)$ during the stretch.
We again took into account four lowest Bloch bands, assumed that initially the population was in the ground Bloch band, and monitored the excitation to 
higher-energy bands. We found that the stretch can be performed rather fast in $\sim 16/E_{R}\sim 730$ $\mu$s, where $E_{R}=3.5$ kHz for $^{87}$Rb,  
while retaining $99.55\%$ of the population in the 
ground Bloch band. The populations of the lowest four bands during the lattice stretch are shown in \fig{fig:Latt-stretch}.

Finally, we can estimate the total time required to generate the 2D cluster state in the blockaded case. The time required for 1D state generation 
$T_{1D}\approx 700$ $\mu$s; 
ramping down the 
$V_{0}$ lattice takes $\approx 250$ $\mu$s and the lattice stretch in the $x$-direction takes $\approx 730$ $\mu$s; the total time is $\approx 1.7$ ms. 
Next, the lattice period in 
the $y$-direction has to be adiabatically reduced, which will similarly take $\approx 730$ $\mu$s, followed by ramping up the $V_{3}$ lattice in $\approx 250$ $\mu$s, and a 
sequence of phase gate operations applied to the $y$-chains, will take $T_{1D}\approx 700$ $\mu$s. Therefore, the total time of the 2D cluster state generation in the blockaded 
case is $T_{2D}\approx 3.4$ ms. The time required for 2D cluster state generation in the no-blockade case can be found summing the time required to produce 
1D states in $x$-direction, $T_{1D}\approx 80$ $\mu$s, the time required to stretch the lattice in $x$-direction and shrink it in $y$ direction 
$\approx 2\times 730$ $\mu$s=$1.46$ ms, followed by application of the phase gate sequence in $y$-direction, requiring time $T_{1D}\approx 80$ $\mu$s. The total 
time needed to produce a 2D cluster state is therefore $T_{2D}\approx 1.62$ ms. These times are much smaller than the qubit coherence times $\sim 1$ s 
\cite{Saffman-error-analysis}, achievable in optical lattices.

\section{Discussion and conclusions}

In Sections II.A.2 and II.B.2 we showed that the phase gate between two neighboring atoms can be realized with an error 
$\epsilon \approx 1.25\cdot 10^{-2}$ and 
$\epsilon \approx 0.21$ without and with dipole blockade in the pair, respectively, if two-photon excitation to Rydberg states is used. 
The major contribution to the error in both cases comes from the 
variation of the two-photon Rabi frequency experienced by each atom due to the spatial variation of the excitation field. This error can be reduced to 
$\epsilon_{\rm \Omega \;var}=(\pi^{2}/2)\left\langle \left(\delta \Omega/\langle \Omega \rangle \right)^{2}\right \rangle$, {\it i.e.} four times, if 
a single-photon excitation is used instead of a two-photon one. Then, $\epsilon_{\rm \Omega \;var} \approx 2.63 \cdot 10^{-3}$ 
and $\epsilon_{\rm \Omega \; var} \approx 3.75 \cdot 10^{-2}$ in the cases without and with dipole blockade, respectively. It will reduce the total gate error to 
$\epsilon \approx 4.3 \cdot 10^{-3}$ 
without and $\epsilon \approx 0.16$ with dipole blockade. 
The error of the blockaded phase gate can be further reduced by increasing the relative qubit energy shift $\Delta_{\rm vec}$ between left and right sites in a double well. 
Namely, the error term $\epsilon_{\rm imp \;block}=(1/2)(\Omega/\Delta_{\rm{vec}})^{2}$ (here one-photon excitation is assumed) can be reduced by increasing 
$\Delta_{\rm vec}$. 
One possible way to increase $\Delta_{\rm{vec}}$ is to use hyperfine states $\ket{F,m_{F}}$ with large $|m_{F}|$ for qubit encoding, 
since $\Delta_{\rm{vec}}\sim m_{F}$. 
For example, if Cs $\ket{F=4,m_{F}=4}$ and $\ket{F=3,m_{F}=3}$ states are used as qubit states $\ket{1}$ and $\ket{0}$, this will allow the increase up 
to $\Delta_{\rm{vec}}\sim 200\;{\rm kHz}\times (m_{F}=4)\sim 800$ kHz. 
The $\Delta_{\rm{vec}}$ can be increased even further 
using atoms with large hyperfine quantum numbers $F$, such as the rare-earth Ho, having $4\le F \le 11$ in the ground $4f^{11}6s^{2}(^{4}I_{15/2})$ state 
\cite{Holmium}, which will allow the increase up to $\Delta_{\rm vec}\sim 1$ MHz. 
Choosing one-photon Rabi frequency $\Omega \sim 100$ kHz, the blockaded gate errors become 
$\epsilon_{\rm imp \; block}=\Omega^{2}/(2\Delta_{\rm{vec}}^{2}) \approx 5\cdot 10^{-3}$ and 
$\epsilon_{\rm Rydb \;decay}=7\pi \gamma/4\Omega \approx 1.75 \cdot 10^{-2}$. The error due to Rabi frequency variation
$\epsilon_{\rm \Omega \; var}=(\pi^{2}/2)\langle \left(\delta \Omega/\Omega\right)^{2} \rangle=(\pi^{2}/4)(ka)^{2}\frac{1-\sqrt{(1+V_{0}/4V_{1})/2}}{1+\sqrt{(1+V_{0}/4V_{1})/2}}$ 
is harder to reduce, since it pretty much depends only on the ratio $V_{0}/V_{1}$, which 
cannot be increased much beyond $V_{0}\sim V_{1}$ in order to have motional states localized in left and right sites of a double-well. Using a very deep 
optical lattice with $V_{0}=200 E_{R}$ and $V_{0}=2V_{1}$, which still results in motional states localized in left and right wells, allows to reduce the 
error to $\epsilon_{\rm \Omega \; var} \approx 1.13 \cdot 10^{-2}$.  
 The transition between 
$\ket{F=4,m_{F}=4}$ and $\ket{F=3,m_{F}=3}$ states of Cs is sensitive to magnetic field at all field values, meaning that the qubit will experience decoherence 
 in these states. To reduce its effect one can keep the qubit in field insensitive $\ket{F=4,m_{F}=0}$ and $\ket{F=3, m_{F}=0}$ states during storage 
and lattice manipulation time and 
transfer it to the $\ket{F=4,m_{F}=4}$, $\ket{F=3,m_{F}=3}$ only when the phase gate is applied. This will result in an additional phase gate 
error due to magnetic field fluctuations $\epsilon_{\rm MF\;fluct} \sim (\delta \omega T_{\rm PG})^{2}$, where $\delta \omega \approx g_{e}\mu_{B} \delta B$ is the 
fluctuation of the qubit transition frequency due to the fluctuating magnetic field $\delta B$ and $g_{e}$ is the electron's gyromagnetic ratio. 
In a recent study \cite{MF-fluctuations} suppression of magnetic field fluctuations down to $\delta B \sim 50$ $\mu$G was demonstrated, resulting 
in $\delta \omega \sim 10^{3}$ s$^{-1}$. One phase gate operation requires $T_{PG}\approx 25$ $\mu$s in our setup, resulting in the dephasing error 
per gate $\epsilon_{\rm MF\;fluct}\sim 6\cdot 10^{-4}$. As a result, the  total gate error in the blockaded case can be reduced to 
$\epsilon \approx 3.44\cdot 10^{-2}$.

At the same time there are errors due to the excitation of atoms in "inactive" latice sites, situated at the minima of the standing wave. With a 
two-photon excitation, the probability to find the pair of "inactive" atoms in the initial two-qubit state after the gate, averaged over all four initial 
states, is $\langle P \rangle \approx 0.75$ and $\langle P \rangle \approx 0.998$ in the cases without and with blockade, respectively. The 
 probability in the former case can be increased using multi-photon, {\it e.g.} four-photon excitation to the Rydberg states. With a four-photon 
exitation, the probability increases up to $\langle P \rangle \approx 0.994$ in the no-blockade case. At the same time the error due to the variation 
of the Rabi frequency increases four times 
up to $\epsilon_{\rm \Omega \; var}=8\pi^{2}\left \langle \left(\delta \Omega/\langle \Omega \rangle \right)^{2}\right \rangle \approx 4.2\cdot 10^{-2}$, giving the total gate error 
$\epsilon \approx 4.4\cdot 10^{-2}$ in the no-blockade case. On the other hand, single-photon excitation which reduces the gate error in the 
blockaded case, leads to the reduced probability to leave atoms in "inactive" sites unaffected $\langle P \rangle \approx 0.87$. 

We can conclude from this analysis that in no-blockade case the multi-photon excitation, {\it e.g.} four-photon, leads to the optimal combination of 
the phase gate error ($\epsilon \approx 4.4\cdot 10^{-2}$) and the error due to the excitation of atoms in "inactive" lattice sites 
($\epsilon_{\rm inact\; exc}=1-\langle P \rangle \approx 6.5\cdot 10^{-3}$). At the 
same time, in the blockaded case one-photon excitation seems to provide the optimal error combination: phase gate error $\epsilon \approx 3.44 \cdot 10^{-2}$ which 
can be achieved in deep optical lattices and with atoms having large vector shifts. It is combined with the probability to have "inactive" atoms 
unaffected $\langle P \rangle \approx 0.87$. We give an error summary for the no-blockade and blockaded cases in Table I.

\begin{table*}
\centering
\caption{Main errors in the no-blockade and blockaded cases.}
\begin{tabular}{c c c}
\hline
  & ${\rm no-blockade}$ &  \\
\hline
${\rm two-photon \; excitation}$ &  & ${\rm four-photon \; excitation}$ \\
\hline
$\epsilon_{\rm \Omega \; var}=(2\pi^{2})\left\langle \left(\delta \Omega/\langle \Omega \rangle \right)^{2}\right \rangle \approx 1.05\cdot 10^{-2}$  &   & $\epsilon_{\rm \Omega \; var}=(8\pi^{2})\left\langle \left(\delta \Omega/\langle \Omega \rangle \right)^{2}\right \rangle \approx 4.2 \cdot 10^{-2}$\\
$\epsilon_{\rm imp \; exc}=1/8\left(V_{\rm int}/\Omega^{2}/\Delta\right)^{2}\approx 1.25 \cdot 10^{-3}$  &  & $\epsilon_{\rm imp \; exc}=1/8\left(V_{\rm int}/\Omega^{4}/\Delta^{3}\right)^{2}\approx 1.25 \cdot 10^{-3}$\\
$\epsilon_{\rm Rydb \; decay1}=2\pi \gamma/V_{\rm int} \approx 6.7\cdot 10^{-4}$ & & $\epsilon_{\rm Rydb \; decay1}=2\pi \gamma/V_{\rm int} \approx 6.7\cdot 10^{-4}$\\
$\epsilon_{\rm Rydb \; decay2}=\pi \gamma/(\Omega^{2}/\Delta)\approx 3\cdot 10^{-5}$ & & $\epsilon_{\rm Rydb \; decay2}=\pi \gamma/(\Omega^{4}/\Delta^{3})\approx 3\cdot 10^{-5}$\\
$\epsilon_{\rm non \; adiab}=(\pi^{2}/4)(a/R)^{2} \approx 2.5 \cdot 10^{-2}$ & & $\epsilon_{\rm non \; adiab}=(\pi^{2}/4)(a/R)^{2} \approx 2.5 \cdot 10^{-2}$\\
$\epsilon \approx 1.25\cdot 10^{-2\;\dagger}$ & & $\epsilon \approx 4.4\cdot 10^{-2\;\dagger}$\\
$\epsilon_{\rm inact \; exc}=1-\langle P \rangle \approx 0.25$ & & $\epsilon_{\rm inact \; exc}=1-\langle P \rangle \approx 6.5 \cdot 10^{-3}$\\
\hline
 & ${\rm with \; blockade}$ & \\
\hline
${\rm two-photon \; excitation}$ &  & ${\rm one-photon \; excitation}$ \\
\hline
$\epsilon_{\rm \Omega \; var}=(2\pi^{2})\left\langle \left(\delta \Omega/\langle \Omega \rangle \right)^{2}\right \rangle \approx 0.15$ & & $\epsilon_{\rm \Omega \; var}=(\pi^{2}/2)\left\langle \left(\delta \Omega/\langle \Omega \rangle \right)^{2}\right \rangle \approx 1.13\cdot 10^{-2}$\\
$\epsilon_{\rm imp \; block}=\left(\Omega^{2}/\Delta\right)^{2}/\left(2|\Delta_{\rm \;vec}|^{2}\right)\approx 2\cdot 10^{-2}$ & & $\epsilon_{\rm imp \; block}=\left(\Omega^{2}/\Delta\right)^{2}/\left(2|\Delta_{\rm \;vec}|^{2}\right)\approx 5\cdot 10^{-3}$\\
$\epsilon_{\rm Rydb \; decay}=7\pi \gamma/(4\Omega^{2}/\Delta)\approx 4.38 \cdot 10^{-2}$ & & $\epsilon_{\rm Rydb \; decay}=7\pi \gamma/(4\Omega)\approx 1.75 \cdot 10^{-2}$\\
$\epsilon_{\rm MF \; fluct}\sim (\delta \omega T_{PG})^{2}\sim 6\cdot 10^{-4}$ & & $\epsilon_{\rm MF \; fluct}\sim (\delta \omega T_{PG})^{2}\sim 6\cdot 10^{-4}$ \\
$\epsilon \approx 0.21^{\dagger}$ & & $\epsilon \approx 3.44 \cdot 10^{-2\;\dagger}$ \\
$\epsilon_{\rm inact \; exc}=1-\langle P \rangle \approx 2\cdot 10^{-3}$ & & $\epsilon_{\rm inact \; exc}=1-\langle P \rangle \approx 0.13$\\
\hline
$^{\dagger}$ not including $\epsilon_{\rm inact \; exc}$ and $\epsilon_{\rm non \; adiab}$ & & \\
\end{tabular}
\label{table:errors}
\end{table*}

It was shown in \cite{fault-tolerance} that fault-tolerant MBQC can be realized with a 3D cluster state using topologically protected gates. In particular, an error threshold 
for an entangling two-qubit gate during the preparation of the cluster state has been calculated as $\epsilon \approx 7.5\cdot 10^{-3}$. Recently the threshold 
was increased to $1.1-1.4\%$ \cite{MBQC-error-thresh-new}. Unfortunately, we cannot directly compare the gate errors in our scheme with this threshold, 
since in \cite{fault-tolerance,MBQC-error-thresh-new} the error was assumed to be partially depolarizing with the phase gate acting as 
$\hat{U}_{PG}=(1-\epsilon)[\hat{I}_{a}\hat{I}_{b}]+(\epsilon/15)([\hat{I}_{a}\hat{X}_{b}]+...+[\hat{Z}_{a}\hat{Z}_{b}])$. In our case the errors are due to leakage out of the computational subspace. 
We note, however, that the errors of the phase gate without and with blockade are higher than the threshold, but not too far from it. 

Our analysis shows that the no-blockade and blockaded case have similar phase gate errors, but it requires a very deep lattice and atoms with large 
hyperfine numbers in the latter case. Additionally, the probabilty to have "inactive" atoms unaffected is higher in the no-blockade compared to the blockaded case. 
On the other hand, the optimal error combination in the no-blockade case requires a four-photon excitation, while in the blockaded case one-photon excitation 
gives the optimal error combination. Generation of a 1D cluster state in the no-blockade case requires almost an order of magnitude less time (80 $\mu$s) compared to the 
blockaded case (700 $\mu$s) due to the lattice change involved in the latter case. The times required to produce a 2D cluster state are, however, 
comparable: 1.62 ms without and 3.4 ms with dipole blockade. From the overall experimental complexity (deep double-well lattice, lattice manipulation 
required) the blockaded scheme seems more difficult to realize.

The same scheme can be applied to generate the cluster state with polar molecules. For example, molecular states with small dipole moments can be used 
to encode a qubit while for the phase gate molecules can be excited to a state with a large dipole moment, such that in this state molecules can interact 
via dipole-dipole interaction. A good candidate is $^{13}$CO \cite{Our-PRA}, where the ground electronic state has a rather small permanent 
dipole moment $\approx 0.1$ D and long-lived nuclear spin sublevels can be used to encode a qubit. It also has a metastable $a\;^{3}\Pi_{0}$ state with 
a significant permanent dipole moment $\approx 1.4$ D, in which molecules can interact via dipole-dipole interaction. Polar molecules, 
that currently or in the near future can be cooled to ultracold temperatures and placed in an optical lattice, are limited to alkali di-atoms, having permanent dipole moments 
of $\mu <6$ Debye ($1$ Debye=$10^{-18}$ esu cm). It limits the dipole-dipole interaction strength $\mu^{2}/R^{3} \sim 100$ kHz between nearest 
neighbors with $R\sim 500$ nm. The rather small interaction strength makes the no-blockade phase gate preferable for polar molecules. As an example, we 
can estimate the gate error for $^{13}$CO. Assuming two-photon 
excitation to the large dipole moment state, the error due to the variation of the Rabi frequency 
$\epsilon_{\rm \Omega \; var}=2\pi^{2}\left \langle \left(\delta \Omega/\langle \Omega \rangle \right)^{2}\right \rangle \approx 1.1\cdot 10^{-2}$, with the lattice depth 
$V_{0}=100E_{R}$. The error due to the finite ratio of the interaction strength to the Rabi frequency $\epsilon_{\rm imp \; exc}=V_{\rm int}^{2}/8(\Omega^{2}/\Delta)^{2}\approx 8\cdot 10^{-5}$ 
assuming $\mu =1.4$ Debye, $R=500$ nm, resulting in $V_{\rm int}\approx 2.5$ kHz and $\Omega^{2}/\Delta \sim 100$ kHz. The large dipole moment molecular state is 
metastable with the lifetime $\sim 500$ ms, giving $\gamma=2$ s$^{-1}$ and the decay-induced errors 
$\epsilon_{\rm Rydb \; decay \;1}=2\pi \gamma/V_{\rm int}\approx 8\cdot 10^{-4}$, 
$\epsilon_{\rm Rydb \; decay \; 2}=\pi \gamma/(\Omega^{2}/\Delta) \approx 10^{-5}$, 
resulting in the total gate error $\epsilon \approx 1.14\cdot 10^{-2}$. As we discussed above for atoms, the gate pulses affect molecules in "inactive" sites 
if two-photon excitation is used. The probability to have the "inactive" molecules unaffected can be increased using multi-photon, {\it e.g.} four-photon 
excitation. In this case the error due to the Rabi frequency variation is four times larger, and the total gate error becomes $\epsilon \approx 4.3 \cdot 10^{-2}$, while 
the probability for "inactive" atoms to stay in the initial qubit states becomes $\langle P \rangle \approx 0.994$. This 
analysis shows that gate errors for molecules and for atoms are comparable.

In conclusion, we propose and analyze generation of a cluster state for measurement-based quantum computing with neutral atoms and polar molecules 
in an optical lattice using van der Waals and 
dipole-dipole interactions. We consider two schemes for implementation of a phase gate between pairs of 
nearest neighbors required to generate the cluster state: 
without and with individual 
addressing within a pair. We show that in the former case the gate error $\epsilon \approx 1.25 \cdot 10^{-2}$ is feasible with two-photon excitation to Rydberg states, 
provided the excitation-deexcitation to interacting states is adiabatic with respect to motional frequency of the lattice. Two-photon excitation, however, 
leads to high error due to excitation of atoms in "inactive" wells at the minima of the standing wave. The optimal combination of the gate error and the 
"inactive" atoms excitation error can be realized using four-photon excitation to Rydberg states, giving the gate error $\epsilon \approx 4.4 \cdot 10^{-2}$ and 
the probability to have "inactive" atoms unaffected $\langle P \rangle \approx 0.994$.  
In the second scheme individual addressing within a pair allows to implement the phase gate using Rydberg (dipole) blockade. Addressing of atoms within a pair 
can be realized in a polarization-gradient double-well lattice. The gate error in this case is
$\epsilon \approx 0.21$, but can be reduced to $\epsilon \approx 3.44\cdot 10^{-2}$ using one-photon excitation to Rydberg states, atoms with large hyperfine quantum numbers, 
such as cesium and holmium, and a very deep optical lattice. With one-photon excitation the probability of "inactive" atoms to stay in their initial state 
is $\langle P \rangle \approx 0.87$.

We also analyze the lattice manipulation required in the latter case to realize the phase gate between all nearest neighbors 
and show under which conditions the manipulation is adiabatic. The total time required to produce a 1D cluster state $T_{1D}\approx 80$ $\mu$s without 
and $T_{1D}\approx 700$ $\mu$s with individual addressing in a pair, where parameters 
of $^{87}$Rb were used for the estimate.

Finally, we show how the scheme for 1D cluster state generation can be extended to realize a universal 2D cluster state and estimate the total time 
required $T_{2D}\approx 1.62$ ms without and $T_{2D}\approx 3.4$ ms with individual addressing in a pair.

\section{Acknowledgments} 

The authors gratefully acknowledge financial support from AFOSR under the MURI award FA9550-09-1-0588.

\section{Appendix A. Calculation of the averaged fidelity of the phase gate without individual addressing}

We calculate the fidelity of the phase gate $\hat{U}_{PG}$ averaged over all initial two-qubit states $\ket{00}$, $\ket{01}$, $\ket{10}$ and $\ket{11}$ as 
follows
\begin{eqnarray}
F=\frac{1}{4}\left[|\langle 00 \vert \hat{U}_{PG}\ket{00}|^{2}+|-\langle 01 \vert \hat{U}_{PG}\ket{01}|^{2}+\right. \nonumber \\
\left. +|-\langle 10 \vert \hat{U}_{PG}\ket{10}|^{2}+|-\langle 11 \vert \hat{U}_{PG}\ket{11}|^{2}\right],\nonumber
\end{eqnarray}
where we project the final state after the phase gate $\hat{U}_{PG}\ket{\epsilon_{1}\epsilon_{2}}$ ($\epsilon_{1,2}=0,1$) on the state, expected after the 
ideal phase gate $\ket{11}\rightarrow -\ket{11}$, $\ket{01}\rightarrow -\ket{01}$, $\ket{10}\rightarrow -\ket{10}$, and $\ket{00}\rightarrow \ket{00}$.
We assume that errors due to different mechanisms are small and can be analyzed separately and added together. We limit our analysis to intrinsic gate 
errors assuming that technical errors can be in principle eliminated.
 
\subsection{Error due to a finite ratio of the interaction strength to the two-photon Rabi frequency $V_{\rm int}/(\Omega^{2}/\Delta)$}

In this subsection we calculate the errors due to imperfect excitation from the $\ket{11}$ to the $\ket{rr}$ state.
In Section IIA we derived a system of equations for the amplitudes $a_{11}$, $a_{+}$ and $a_{rr}$ of the two-qubit states $\ket{11}$, $\ket{+}=(\ket{1r}+\ket{r1})/\sqrt{2}$, 
 and $\ket{rr}$ in the case when the initial state is $\ket{11}$ (see Eq.(\ref{eq:eqs_11})). 
The corresponding Hamiltonian is
\begin{equation} 
H/\hbar=
\left(\begin{array}{ccc}
-\frac{2\Omega_{1}^{2}}{\Delta} & -\frac{\sqrt{2}\Omega_{1}\Omega_{2}}{\Delta} & 0 \\
-\frac{\sqrt{2}\Omega_{1}\Omega_{2}}{\Delta} & -\frac{\Omega_{1}^{2}+\Omega_{2}^{2}}{\Delta} & -\frac{\sqrt{2}\Omega_{1}\Omega_{2}}{\Delta} \\
0 & -\frac{\sqrt{2}\Omega_{1}\Omega_{2}}{\Delta} & V_{\rm int}-\frac{2\Omega_{1}^{2}}{\Delta}
\end{array} \right)
\end{equation} 

Assuming $\Omega_{1}=\Omega_{2}=\Omega$ for the simplicity of the analysis and $V_{\rm int} \ll \Omega^{2}/\Delta$ we can calculate the eigenvalues of the Hamiltonian:

\begin{eqnarray}
\lambda_{1}=-\frac{2\Omega^{2}}{\Delta}+\frac{V_{\rm int}}{2}-\frac{1}{32}\frac{V_{\rm int}^{3}}{(\Omega^{2}/\Delta)^{2}},\nonumber \\
\lambda_{2}=\frac{1}{4}V_{\rm int}+\frac{5}{64}\frac{V_{\rm int}^{2}}{\Omega^{2}/\Delta},\nonumber \\
\lambda_{3}=-\frac{4\Omega^{2}}{\Delta}+\frac{1}{4}V_{\rm int}-\frac{5}{64}\frac{V_{\rm int}^{2}}{\Omega^{2}/\Delta}, \nonumber
\end{eqnarray}
with the corresponding eigenfunctions

\begin{eqnarray}
\ket{\Psi_{1}}=\frac{1}{\sqrt{2}}\left(1+\frac{1}{32}\left(\frac{V_{\rm int}}{\Omega^{2}/\Delta}\right)^{2}\right)\ket{11}-\frac{V_{\rm int}}{4\Omega^{2}/\Delta}\ket{+} \nonumber \\
-\frac{1}{\sqrt{2}}\left(1-\frac{3}{32}\left(\frac{V_{\rm int}}{\Omega^{2}/\Delta}\right)^{2}\right)\ket{rr},\nonumber \\
\ket{\Psi_{2}}=-\frac{1}{2}\left(1-\frac{3}{16}\frac{V_{\rm int}}{\Omega^{2}/\Delta}-\frac{25}{512}\frac{V_{\rm int}^{2}}{(\Omega^{2}/\Delta)^{2}}\right)\ket{11}+ \nonumber \\
+\frac{1}{\sqrt{2}}\left(1-\frac{1}{16}\frac{V_{\rm int}}{\Omega^{2}/\Delta}-\frac{17}{512}\frac{V_{\rm int}^{2}}{(\Omega^{2}/\Delta)^{2}}\right)\ket{+} \nonumber \\
-\frac{1}{2}\left(1+\frac{5}{16}\frac{V_{\rm int}}{\Omega^{2}/\Delta}+\frac{23}{512}\frac{V_{\rm int}^{2}}{(\Omega^{2}/\Delta)^{2}}\right)\ket{rr}, \nonumber \\
\ket{\Psi_{3}}=\frac{1}{2}\left(1+\frac{3}{16}\frac{V_{\rm int}}{\Omega^{2}/\Delta}-\frac{25}{512}\frac{V_{\rm int}^{2}}{(\Omega^{2}/\Delta)^{2}}\right)\ket{11}+ \nonumber \\
+\frac{1}{\sqrt{2}}\left(1+\frac{1}{16}\frac{V_{\rm int}}{\Omega^{2}/\Delta}-\frac{17}{512}\frac{V_{\rm int}^{2}}{(\Omega^{2}/\Delta)^{2}}\right)\ket{+}+ \nonumber \\
+\frac{1}{2}\left(1-\frac{5}{16}\frac{V_{\rm int}}{\Omega^{2}/\Delta}+\frac{23}{512}\frac{V_{\rm int}^{2}}{(\Omega^{2}/\Delta)^{2}}\right)\ket{rr} \nonumber
\end{eqnarray}
up to the second order in $V_{\rm int}/(\Omega^{2}/\Delta)$. 

The state $\ket{11}$ evolves as 
\begin{eqnarray}
\ket{\Psi}=a_{1}\ket{\Psi_{1}}e^{-i\lambda_{1}t}+a_{2}\ket{\Psi_{2}}e^{-i\lambda_{2}t}+a_{3}\ket{\Psi_{3}}e^{-i\lambda_{3}t},\nonumber
\end{eqnarray}
where the coefficients $a_{i}$ are determined from the condition $\ket{\Psi(t=0)}=\ket{11}$ and are given by:
\begin{eqnarray}
a_{1}=\frac{1}{\sqrt{2}}\left(1+\frac{1}{32}\left(\frac{V_{\rm int}}{\Omega^{2}/\Delta}\right)^{2}\right), \nonumber \\
a_{2}=-\frac{1}{2}\left(1-\frac{3}{16}\frac{V_{\rm int}}{\Omega^{2}/\Delta}-\frac{25}{512}\left(\frac{V_{\rm int}}{\Omega^{2}/\Delta}\right)^{2}\right), \nonumber \\
a_{3}=\frac{1}{2}\left(1+\frac{3}{16}\frac{V_{\rm int}}{\Omega^{2}/\Delta}-\frac{25}{512}\left(\frac{V_{\rm int}}{\Omega^{2}/\Delta}\right)^{2}\right). \nonumber
\end{eqnarray} 
The phase gate is realized in three steps: first, a $\pi$ pulse is applied to both atoms during the time $T$ such that $\Omega^{2}T/\Delta=\pi/2$. Next, atoms interact 
in the $\ket{rr}$ state and accumulate the phase $V_{\rm int}T_{\rm int}=\pi$. Finally, a second $\pi$ pulse of the same duration $T$ de-excites atoms back 
to the $-\ket{11}$ state. The wavefunction at the end of the gate is as follows
\begin{eqnarray*}
\ket{\Psi} & = & a_{1}\ket{\Psi_{1}}e^{-i\frac{\pi}{2}\frac{V_{\rm int}}{\Omega^{2}/\Delta}}+a_{2}\ket{\Psi_{2}}e^{-i\frac{\pi}{4}\frac{V_{\rm int}}{\Omega^{2}/\Delta}-\frac{5\pi i}{64}\left(\frac{V_{\rm int}}{\Omega^{2}/\Delta}\right)^{2}}+\nonumber \\
&& +a_{3}\ket{\Psi_{3}}e^{-i\frac{\pi}{4}\frac{V_{\rm int}}{\Omega^{2}/\Delta}+\frac{5\pi i}{64}\left(\frac{V_{\rm int}}{\Omega^{2}/\Delta}\right)^{2}}+\nonumber \\
&& +2\left[\frac{1}{2}\left(1-\frac{1}{16}\left(\frac{V_{\rm int}}{\Omega^{2}/\Delta}\right)^{2}\right)e^{-i\frac{\pi}{4}\frac{V_{\rm int}}{\Omega^{2}/\Delta}}+\right. \nonumber \\
&& \left. +\frac{1}{4}\left(1+\frac{1}{8}\frac{V_{\rm int}}{\Omega^{2}/\Delta}-\frac{1}{16}\left(\frac{V_{\rm int}}{\Omega^{2}/\Delta}\right)^{2}\right) \times \right. \nonumber \\
&&\left. \times e^{-i\frac{\pi}{8}\frac{V_{\rm int}}{\Omega^{2}/\Delta}-\frac{5\pi i}{128}\left(\frac{V_{\rm int}}{\Omega^{2}/\Delta}\right)^{2}}+\right. \nonumber \\
&& \left. +\frac{1}{4}\left(1-\frac{1}{8}\frac{V_{\rm int}}{\Omega^{2}/\Delta}-\frac{1}{16}\left(\frac{V_{\rm int}}{\Omega^{2}/\Delta}\right)^{2}\right)\times \right. \nonumber \\
&&\left. \times e^{-i\frac{\pi}{8}\frac{V_{\rm int}}{\Omega^{2}/\Delta}+\frac{5\pi i}{128}\left(\frac{V_{\rm int}}{\Omega^{2}/\Delta}\right)^{2}}\right]\times \nonumber \\
&& \times\left(-\frac{1}{\sqrt{2}}\left(1-\frac{3}{32}\left(\frac{V_{\rm int}}{\Omega^{2}/\Delta}\right)^{2}\right)\ket{\Psi_{1}}e^{-i\frac{\pi}{4}\frac{V_{\rm int}}{\Omega^{2}/\Delta}}+\right. \nonumber \\
&& \left. +\frac{1}{\sqrt{2}}\left(1+\frac{5}{16}\frac{V_{\rm int}}{\Omega^{2}/\Delta}+\frac{23}{512}\left(\frac{V_{\rm int}}{\Omega^{2}/\Delta}\right)^{2}\right)\ket{\Psi_{2}}\times \right. \nonumber \\
&& \left. \times e^{-i\frac{\pi}{8}\frac{V_{\rm int}}{\Omega^{2}/\Delta}-\frac{5\pi i}{128}\left(\frac{V_{\rm int}}{\Omega^{2}/\Delta}\right)^{2}} \right. \nonumber \\
&& \left. -\frac{1}{\sqrt{2}}\left(1-\frac{5}{16}\frac{V_{\rm int}}{\Omega^{2}/\Delta}+\frac{23}{512}\left(\frac{V_{\rm int}}{\Omega^{2}/\Delta}\right)^{2}\right)\ket{\Psi_{3}}\times \right. \nonumber \\
&& \left. \times e^{-i\frac{\pi}{8}\frac{V_{\rm int}}{\Omega^{2}/\Delta}+\frac{5\pi i}{128}\left(\frac{V_{\rm int}}{\Omega^{2}/\Delta}\right)^{2}}\right).
\end{eqnarray*}

Expanding the exponents in $V_{\rm int}/(\Omega^{2}/\Delta)$ and keeping terms up to the second order we get 
\begin{eqnarray}
\langle 11 \vert \Psi\rangle=-1+\left(\frac{1}{4}+\frac{9\pi^{2}}{128}\right)\left(\frac{V_{\rm int}}{\Omega^{2}/\Delta}\right)^{2}+\frac{3\pi i}{8}\frac{V_{\rm int}}{\Omega^{2}/\Delta}. \nonumber
\end{eqnarray}
As a result, $\left|-\langle 11 \vert \Psi\rangle\right|^{2}=1-\frac{V_{\rm int}^{2}}{2(\Omega^{2}/\Delta)^{2}}$, with the corresponding error $\frac{V_{\rm int}^{2}}{2(\Omega^{2}/\Delta)^{2}}$. 
This error is present only for the initial state $\ket{11}$, since for the initial states $\ket{01}$, $\ket{10}$ and $\ket{00}$ as is seen from Eqs.(3),(5) 
the state $\ket{rr}$ is not populated and no interaction is involved. The averaged over all initial states error is therefore $\frac{V_{\rm int}^{2}}{8(\Omega^{2}/\Delta)^{2}}$.

This analysis also allows to calculate the error due to the unwanted interaction of atoms belonging to different excited pairs. If we assume that 
$V_{\rm int}T_{\rm int}=\pi(1+\delta V_{\rm int}/V_{\rm int})$, where $\delta V_{\rm int}$ is due to the interaction between closest pairs, the corresponding 
error averaged over all initial states is $(3\pi^{2}/16)\left(1/8+19\pi^{2}/256\right)(\delta V_{\rm int}/V_{\rm int})(V_{\rm int}/(\Omega^{2}/\Delta))^{3}$.

\subsection{Error due to decay of Rydberg states}

In this subsection we calculate the error due to decay of Rydberg 
states.
We can calculate this error using Eq.(\ref{eq:wavefunct_11}) for the $\ket{11}$ initial state assuming that the $\ket{r}$ state decays 
with the rate $\gamma$. Then, the $\ket{11}$ state at the end of the gate becomes
\begin{eqnarray}
\ket{\Psi}=-\ket{11}e^{-\gamma \pi/(\Omega^{2}/\Delta) -2\gamma \pi/V_{\rm int}}\approx \nonumber \\
\approx -\ket{11}\left(1-\frac{\gamma \pi}{\Omega^{2}/\Delta}-\frac{2\gamma \pi}{V_{\rm int}}\right) \nonumber,
\end{eqnarray}
where we expanded the exponent using the smallness of $\gamma/(\Omega^{2}/\Delta)$, $\gamma/V_{\rm int}$.
As a result, $\left|-\langle 11 \vert \Psi\rangle\right|^{2}=1-\frac{2\gamma \pi}{\Omega^{2}/\Delta}-\frac{4\gamma \pi}{V_{\rm int}}$, giving the error 
$\frac{2\gamma \pi}{\Omega^{2}/\Delta}+\frac{4\gamma \pi}{V_{\rm int}}$.

If the initial state is $\ket{01}$ the system evolves according to Eq.(\ref{eq:wavefunct_01}). At the end of the gate 
$\ket{\Psi}= -\ket{01}e^{-\gamma \pi/(2\Omega^{2}/\Delta)-\gamma \pi/V_{\rm int}}$. The overlap with the desired $-\ket{01}$ state is then 
$\left|-\langle 01 \vert \Psi\rangle \right|^{2}=e^{-\gamma \pi/(\Omega^{2}/\Delta)-2\gamma \pi/V_{\rm int}}$, resulting in the error $\gamma \pi/(\Omega^{2}/\Delta)+2\gamma \pi/V_{\rm int}$. 
Similar result is obtained for the $\ket{10}$ state. There is no error due to the Rydberg state decay for the $\ket{00}$ initial state. The averaged over all 
initial states error is according to Eq.(\ref{eq:fidelity}) given by $\gamma \pi/(\Omega^{2}/\Delta)+2\gamma \pi/V_{\rm int}$.

\subsection{Error due to non-adiabatic excitation to the $\ket{rr}$ state}

When two atoms in the $\ket{11}$ internal state and in the ground motional states of the lattice $\ket{g_{1}g_{2}}$ are transferred to a doubly excited 
Rydberg state $\ket{rr}$, the state $\ket{rr}\otimes \ket{g_{1}g_{2}}$ is no longer the eigenstate of the total Hamiltonian, including vdW or 
dipole-dipole interaction. The interaction admixes higher-energy motional states to the original ground state as well as other Rydberg states to 
a much smaller extent. As a result, new eigenstates are superpositions of several motional states. Let us call the new ground motional state of two 
atoms in the presence of vdW or dipole-dipole interactions $\ket{\tilde{g}}$, while $\ket{\tilde{e}_{k}}$ are the higher-energy motional states. If the 
interaction is not very strong only closest in energy states are admixed, $\ket{\tilde{g}}\approx \ket{g_{1}g_{2}}+\alpha\left(\ket{g_{1}e_{2}^{(1)}}+\ket{e_{1}^{(1)}g_{2}}\right)$ in the first order perturbation theory, 
where $\ket{e_{i}^{(j)}}$ is the j$^{\rm th}$ excited motional state of the i$^{\rm th}$ atom, and $\alpha=\langle g_{1}g_{2} \vert \hat{V}_{\rm int}\ket{g_{1}e_{2}^{(1)}}/(E_{g_{1}g_{2}}-E_{g_{1}}-E_{e_{2}^{(1)}})$ 
(assuming $\langle g_{1}g_{2} \vert \hat{V}_{\rm int}\ket{g_{1}e_{2}^{(1)}}=\langle g_{1}g_{2} \vert \hat{V}_{\rm int}\ket{e_{1}^{(1)}g_{2}}$). If the atoms are excited to $\ket{rr}$ adiabatically 
with respect to the energy separation between $\ket{\tilde{g}}$ and higher-energy states (which is of the order of the motional energy splitting), the 
state $\ket{g_{1}g_{2}}$ gradually evolves into $\ket{\tilde{g}}$ and there is no uncertainty in the interaction strength. If, however, the excitation 
is non-adiabatic, the state $\ket{g_{1}g_{2}}$ does not change during the excitation, {\it i.e.} it is now a superposition of the new eigenstates 
$\ket{g_{1}g_{2}}\approx \ket{\tilde{g}}+\beta\ket{\tilde{e}_{1}}$, where $\beta\approx -\alpha$. As a result, after the first $\pi$ pulse 
\begin{eqnarray}
\ket{11}\otimes \ket{g_{1}g_{2}}\rightarrow \ket{rr}\otimes \ket{g_{1}g_{2}}=\ket{rr}\otimes (\ket{\tilde{g}}+\beta \ket{\tilde{e}_{1}}). \nonumber 
\end{eqnarray}
During the interaction time the state $\ket{\tilde{e}_{1}}$ acquires a phase factor $\exp(-i\Delta E T_{\rm int}/\hbar)$, where $\Delta E$ is the energy 
difference between the $\ket{\tilde{e}_{1}}$ and $\ket{\tilde{g}}$ states. After the interaction time the system is in the state 
\begin{eqnarray}
\ket{rr}\otimes (\ket{\tilde{g}}+\beta e^{-i\pi \Delta E/V_{\rm int}\hbar}\ket{\tilde{e}_{1}}). \nonumber 
\end{eqnarray}
The second $\pi$ pulse brings the system back to the $\ket{11}$ state without changing the motional state. As a result, after the second $\pi$ pulse the 
state looks as follows
\begin{eqnarray}
-\ket{11}\otimes (\ket{\tilde{g}}+\beta e^{-i\pi \Delta E/V_{\rm int}\hbar}\ket{\tilde{e}_{1}})= \nonumber \\
=-\ket{11}\otimes (\ket{g_{1}g_{2}}+\beta(e^{-i\pi \Delta E/V_{\rm int}\hbar}-1)\ket{\tilde{e}_{1}}). \nonumber
\end{eqnarray}
The projection to the ideal $-\ket{11}\otimes \ket{g_{1}g_{2}}$ is then given by $1+|\beta|^{2}(e^{-i\pi \Delta E/V_{\rm int}\hbar}-1)\approx 1+|\alpha|^{2}(-i\pi \Delta E/V_{\rm int}\hbar-(\pi^{2}/2)(\Delta E/V_{\rm int}\hbar)^{2})$. 
Since $\Delta E \approx \hbar \omega$, and $\langle g_{1}g_{2} \vert \hat{V}_{\rm int}\ket{g_{1}e_{2}^{(1)}}\sim (a/R)\langle g_{1}g_{2} \vert \hat{V}_{\rm int}\ket{g_{1}g_{2}}=(a/R)V_{\rm int}$, 
the perturbation theory gives $|\alpha|\sim (a/R)(V_{\rm int}/\omega)$. As a result, the projection to the ideal final state is $1+(a/R)^{2}(V_{\rm int}/\omega)^{2}\left(-i\pi \omega/V_{\rm int}-\pi^{2}\omega^{2}/2V_{\rm int}^{2}\right)$, 
which gives the error $\pi^{2}(a/R)^{2}-\pi^{2}(a/R)^{4}(V_{\rm int}/\omega)^{2}$. Since this error is present only for the $\ket{11}$ initial state the 
averaged error is $(\pi^{2}/4)(a/R)^{2}-(\pi^{2}/4)(a/R)^{4}(V_{\rm int}/\omega)^{2}$.

\subsection{Error due to the variation of the Rabi frequency}

In this subsection we calculate the error caused by the variation of the Rabi frequency experienced by each atom 
due to the finite width of the ground motional state. 
If the initial state is $\ket{11}$, the effect of the Rabi frequency spread can be calculated from Eq.(\ref{eq:wavefunct_11}), where the Rabi frequency 
$\Omega=\langle \Omega \rangle +\delta \Omega$, and we again assume $\Omega_{1}=\Omega_{2}=\Omega$. Assuming also that for the two 
$\pi$ pulses $\langle \Omega \rangle ^{2}T/\Delta=\pi/2$, the state at the end of the gate is 
given by
\begin{eqnarray}
\ket{\Psi} & = & e^{4\pi i\delta \Omega/\langle \Omega \rangle+2\pi i \left(\delta \Omega/\langle \Omega \rangle \right)^{2}}\left[\frac{\ket{11}-\ket{rr}}{2}+\right. \nonumber \\
&&  +\frac{\ket{11}-\sqrt{2}\ket{+}+\ket{rr}}{4}e^{-4\pi i\delta \Omega/\langle \Omega \rangle-2\pi i \left(\delta \Omega/\langle \Omega \rangle \right)^{2}}+ \nonumber \\
&&  +\frac{\ket{11}+\sqrt{2}\ket{+}+\ket{rr}}{4}e^{4\pi i\delta \Omega/\langle \Omega \rangle+2\pi i \left(\delta \Omega/\langle \Omega \rangle \right)^{2}}+ \nonumber \\
&& +\left(1+\frac{1}{2}e^{-2\pi i\delta \Omega/\langle \Omega \rangle-\pi i \left(\delta \Omega/\langle \Omega \rangle \right)^{2}}+ \right. \nonumber \\
&& \left. + \frac{1}{2}e^{2\pi i\delta \Omega/\langle \Omega \rangle+\pi i \left(\delta \Omega/\langle \Omega \rangle \right)^{2}}\right)\times  \nonumber \\
&&  \times\left(-\frac{\ket{11}-\ket{rr}}{2}-\frac{\ket{11}-\sqrt{2}\ket{+}+\ket{rr}}{4}\times  \right. \nonumber \\
&& \left. \times e^{-2\pi i\delta \Omega/\langle \Omega \rangle-\pi i \left(\delta \Omega/\langle \Omega \rangle \right)^{2}} \right. \nonumber \\
&& \left. \left. -\frac{\ket{11}+\sqrt{2}\ket{+}+\ket{rr}}{4}e^{2\pi i\delta \Omega/\langle \Omega \rangle+\pi i \left(\delta \Omega/\langle \Omega \rangle \right)^{2}}\right)\right] \nonumber
\end{eqnarray}
Expanding the exponents up to the second order in $\delta \Omega/\langle \Omega \rangle$ we have $\ket{\Psi}=-\ket{11}e^{4\pi i \delta \Omega/\langle \Omega \rangle+2\pi i\left(\delta \Omega/\langle \Omega \rangle \right)^{2}}$, 
which gives the projection to the desired $-\ket{11}$ state $\left|-\langle 11 \vert \Psi \rangle \right|^{2}\sim O\left((\delta \Omega/\langle \Omega \rangle)^{3}\right)$.

If the initial state is $\ket{01}$ the wavefunction is given by Eq.(\ref{eq:wavefunct_01})
\begin{eqnarray}
\ket{\Psi} & = & e^{i\Omega^{2}(\frac{1}{\Delta}+\frac{1}{\Delta+\Delta_{hf}})t}\left(\frac{\ket{01}-\ket{0r}}{2}e^{-i\Omega^{2}t/\Delta}+\right. \nonumber \\
&& \left. +\frac{\ket{01}+\ket{0r}}{2}e^{i\Omega^{2}t/\Delta}\right). \nonumber 
\end{eqnarray}
Assuming again $\Omega=\langle \Omega \rangle +\delta \Omega$ and $\langle \Omega \rangle^{2}T/\Delta=\pi$ after the two $\pi$ pulses, the 
wavefunction becomes
\begin{eqnarray}
\ket{\Psi} & = & -e^{4\pi i\delta \Omega/\langle \Omega \rangle +2\pi i (\delta \Omega/\langle \Omega \rangle)^{2}-i\pi \Delta_{hf}/\Delta}\times \nonumber \\
&&\times \left(\ket{01}\cos\left(2\pi \delta \Omega/\langle \Omega \rangle+\pi (\delta \Omega/\langle \Omega \rangle)^{2}\right)+\right. \nonumber \\
&& \left. +i\ket{0r}\sin \left(2\pi \delta \Omega/\langle \Omega \rangle+\pi (\delta \Omega/\langle \Omega \rangle)^{2}\right)\right).\nonumber 
\end{eqnarray}

The projection to the ideal $-\ket{01}$ state is then
\begin{eqnarray}
-\langle 01 \vert \Psi\rangle & = & e^{-i\pi \Delta_{hf}/\Delta+4\pi i \delta \Omega/\langle \Omega \rangle +2\pi i(\delta \Omega/\langle \Omega \rangle)^{2}}\times \nonumber \\
&& \times \left(1-2\pi^{2}\left(\frac{\delta \Omega}{\langle \Omega \rangle}\right)^{2}\right),\nonumber  
\end{eqnarray}
giving the error $4\pi^{2}(\delta \Omega/\langle \Omega \rangle)^{2}$. A similar result is obtained for the $\ket{10}$ state.  

Finally, if the initial state is $\ket{00}$, it evolves according to Eq.(\ref{eq:eqs_00}). After the phase gate the state turns into 
$\ket{\Psi}=\ket{00}e^{2\pi i \delta \Omega/\langle \Omega \rangle +i\pi (\delta \Omega/\langle \Omega \rangle)^{2}}$. The overlap with the desired 
$\ket{00}$ state is $\left|\langle 00 \vert \Psi\rangle \right|^{2}=1$.

The averaged over all initial states according to Eq.(\ref{eq:fidelity}) error is then $2\pi^{2}\left \langle \left(\delta \Omega^{2}/\langle \Omega \rangle \right)^{2}\right \rangle$, where the 
averaging in the error expression is over the ground motional state wavefunction.

\subsection{Excitation of atoms in minima of standing wave excitation pulse}

We also need to estimate the probability that the atoms in "inactive" lattice sites, {\it i.e.} at the minima of the standing wave excitation pulse, are not 
affected. Let us denote $\tilde{\Omega}^{2}/\Delta$ the two-photon Rabi frequency at these sites. Now, as can be seen from Eq.(\ref{eq:wavefunct_11}) the $\ket{11}$ initial state after the gate becomes
\begin{eqnarray}
\ket{\Psi}=e^{2\pi i \frac{\tilde{\Omega}^{2}}{\Omega^{2}}}\left[\frac{\ket{11}}{4}\left(\cos\left(2\pi\frac{\tilde{\Omega}^{2}}{\Omega^{2}}\right) 
+4\cos\left(\pi \frac{\tilde{\Omega}^{2}}{\Omega^{2}}\right) -1 \right)+ \right. \nonumber \\
 +\frac{\ket{rr}}{4}\left(\cos\left(2\pi\frac{\tilde{\Omega}^{2}}{\Omega^{2}}\right) -1 \right)+  \nonumber \\
\left. +\frac{i\sqrt{2}\ket{+}}{4}\left(\sin\left(2\pi\frac{\tilde{\Omega}^{2}}{\Omega^{2}}\right)+\sin\left(\pi\frac{\tilde{\Omega}^{2}}{\Omega^{2}}\right) \right)\right]. \nonumber
\end{eqnarray}
The overlap with the initial $\ket{11}$ state is then
\begin{eqnarray}
\langle 11 \vert \Psi \rangle=\frac{e^{2\pi i \tilde{\Omega}^{2}/\Omega^{2}}}{2}\left(\cos^{2}\left(\pi \tilde{\Omega}^{2}/\Omega^{2}\right)-1+2\cos\left(\pi \tilde{\Omega}^{2}/\Omega^{2}\right)\right]. \nonumber
\end{eqnarray}

The initial $\ket{01}$ state (the same for $\ket{10}$ state), according to Eq.(\ref{eq:wavefunct_01}) after the gate becomes 
\begin{eqnarray}
\ket{\Psi}=e^{2\pi i \frac{\tilde{\Omega}^{2}}{\Omega^{2}}}\left(\ket{01}\cos\left(\pi \frac{\tilde{\Omega}^{2}}{\Omega^{2}}\right)
+i\ket{0r}\sin\left(\pi \frac{\tilde{\Omega}^{2}}{\Omega^{2}}\right)\right), \nonumber
\end{eqnarray}
which results in the overlap with the initial $\ket{01}$ state $\langle 01 \vert \Psi \rangle=e^{2\pi i \tilde{\Omega}^{2}/\Omega^{2}}\cos\left(\pi \tilde{\Omega}^{2}/\Omega^{2}\right)$.

The initial state $\ket{00}$ becomes $\ket{\Psi}=\ket{00}e^{2i\pi \tilde{\Omega}^{2}/\Omega^{2}}$ after the gate, according to Eq.(\ref{eq:eqs_00}).

The Rabi frequency at the minima of the standing wave $\tilde{\Omega}=\Omega_{0}\cos(3\pi/8)$, at the maxima $\Omega=\Omega_{0}\cos(7\pi/8)$, as a result, 
the ratio $\tilde{\Omega}^{2}/\Omega^{2}\approx 0.17$. The probability to find the pair of atoms in the initial state after the gate, averaged over all 
initial states, is then $\langle P \rangle =(1/4)\left(|\langle 11 \vert \Psi \rangle|^{2}+|\langle 01 \vert \Psi \rangle|^{2}+|\langle 10 \vert \Psi \rangle|^{2}+|\langle 00 \vert \Psi \rangle|^{2}\right)\approx 0.75$.

\section{Appendix B. Calculation of the averaged fidelity of the phase gate with individual addressing (blockaded case)}

Next,we calculate the errors of the phase gate with individual addressing in a pair of atoms. We start by calculating the error due to the finite ratio of the two-photon 
Rabi frequency to the vector shift $\Delta_{\rm vec}$ between atoms in the left and right wells and between the Rabi frequency and the interaction 
strength $V_{\rm int}$, resulting in imperfect individual addressing. 

\subsection{Error due to a finite ratio of the Rabi frequency to the vector shift $(\Omega^{2}/\Delta)/\Delta_{\rm vec}$ and interaction strength 
$(\Omega^{2}/\Delta)/V_{\rm int}$}

\subsubsection{$\ket{11}$ initial state}

First we consider the $\ket{11}$ initial state. The phase gate is implemented in the following way: a $\pi$ pulse resonant to the $\ket{1}-\ket{r}$ 
transition of the control, {\it e.g.} the left atom is applied. The states $\ket{11}$, $\ket{1r}$, $\ket{r1}$ and $\ket{rr}$ (other states are far detuned and have 
much smaller amplitudes) evolve according to the Hamiltonian:

\begin{widetext}
\begin{equation} 
H/\hbar=
\left(\begin{array}{cccc}
-\left(\frac{\Omega_{1}^{2}}{\Delta+\Delta_{\rm vec}}+\frac{\Omega_{2}^{2}}{\Delta}\right) & -\frac{\Omega_{1}\Omega_{2}}{\Delta+\Delta_{\rm vec}} & -\frac{\Omega_{1}\Omega_{2}}{\Delta} & 0 \\
-\frac{\Omega_{1}\Omega_{2}}{\Delta+\Delta_{\rm vec}} & \left(\Delta_{\rm vec}-\frac{\Omega_{1}^{2}}{\Delta+\Delta_{\rm vec}}-\frac{\Omega_{2}^{2}}{\Delta+\Delta_{\rm vec}}\right) & 0 & -\frac{\Omega_{1}\Omega_{2}}{\Delta+\Delta_{\rm vec}} \\
-\frac{\Omega_{1}\Omega_{2}}{\Delta} & 0 & -\left(\frac{\Omega_{1}^{2}}{\Delta+\Delta_{\rm vec}}+\frac{\Omega_{2}^{2}}{\Delta}\right) & -\frac{\Omega_{1}\Omega_{2}}{\Delta+\Delta_{\rm vec}} \\
0 & -\frac{\Omega_{1}\Omega_{2}}{\Delta+\Delta_{\rm vec}} & -\frac{\Omega_{1}\Omega_{2}}{\Delta+\Delta_{\rm vec}} & \left(V_{\rm int}+\Delta_{\rm vec}-\frac{\Omega_{2}^{2}}{\Delta+\Delta_{\rm vec}}-\frac{\Omega_{2}^{2}}{\Delta+\Delta_{\rm vec}}\right)
\end{array} \right)
\label{eq:eqs_ind_addr_11_1}
\end{equation} 
\end{widetext}
We assume $\Omega_{1}=\Omega_{2}=\Omega$,  $V_{\rm int}\gg \Delta_{\rm vec} \gg \Omega^{2}/\Delta$, and $\Delta \gg \Delta_{\rm vec},\;V_{\rm int}$. 
In the analysis we will, therefore, keep terms up to the first order in $(\Omega^{2}/\Delta)/V_{\rm int}$ and up to the second order in $(\Omega^{2}/\Delta)/\Delta_{\rm vec}$. 
Neglecting $a_{1r}\sim (\Omega^{2}/\Delta \Delta_{\rm vec})a_{11}$ and $a_{rr}\sim (\Omega^{2}/\Delta V_{\rm int})a_{r1}$ as small, we have a system of equations 
for $a_{11}$ and $a_{r1}$ governed by the Hamiltonian:
\begin{widetext}
\begin{equation}
H/\hbar=
\left(\begin{array}{cc}
-\left(\frac{2\Omega^{2}}{\Delta}+\frac{(\Omega^{2}/\Delta)^{2}}{\Delta_{\rm vec}-2\Omega^{2}/\Delta}\right) & -\frac{\Omega^{2}}{\Delta}\left(1+\frac{(\Omega^{2}/\Delta)^{2}}{V_{\rm int}(\Delta_{\rm vec}-2\Omega^{2}/\Delta)}\right)\\
-\frac{\Omega^{2}}{\Delta}\left(1+\frac{(\Omega^{2}/\Delta)^{2}}{V_{\rm int}(\Delta_{\rm vec}-2\Omega^{2}/\Delta)}\right) & -\left(\frac{2\Omega^{2}}{\Delta}+\frac{(\Omega^{2}/\Delta)^{2}}{V_{\rm int}}\right)
\end{array} \right)
\label{eq:eqs_ind_addr_11_2}
\end{equation}
\end{widetext}
The corresponding eigenstates and eigenfunctions are
\begin{eqnarray}
\lambda_{1,2} & = & -\frac{2\Omega^{2}}{\Delta}-\frac{(\Omega^{2}/\Delta)^{2}}{2\Delta_{\rm vec}}\left(1+\frac{2\Omega^{2}/\Delta}{\Delta_{\rm vec}}\right)\nonumber \\
&& -\frac{(\Omega^{2}/\Delta)^{2}}{2\Delta_{\rm vec}}\pm \frac{\Omega^{2}}{\Delta}\left(1+\frac{(\Omega^{2}/\Delta)^{2}}{8\Delta_{\rm vec}^{2}}\right)
\end{eqnarray}
and 
\begin{eqnarray}
\label{eq:wavefunct_ind_addr_11}
\ket{\Psi_{1}} & = & -\frac{1}{\sqrt{2}}\left(1-\frac{\Omega^{2}/\Delta}{4\Delta_{\rm vec}}+\frac{\Omega^{2}/\Delta}{4V_{\rm int}}-\frac{17(\Omega^{2}/\Delta)^{2}}{32\Delta_{\rm vec}^{2}}\right)\ket{11}+\nonumber \\
&& +\frac{1}{\sqrt{2}}\left(1+\frac{\Omega^{2}/\Delta}{4\Delta_{\rm vec}}-\frac{\Omega^{2}/\Delta}{4V_{\rm int}}+\frac{15(\Omega^{2}/\Delta)^{2}}{32\Delta_{\rm vec}^{2}}\right)\ket{r1}, \nonumber \\
\ket{\Psi_{2}} & = & \frac{1}{\sqrt{2}}\left(1+\frac{\Omega^{2}/\Delta}{4\Delta_{\rm vec}}-\frac{\Omega^{2}/\Delta}{4V_{\rm int}}+\frac{15(\Omega^{2}/\Delta)^{2}}{32\Delta_{\rm vec}^{2}}\right)\ket{11}+\nonumber \\
&& +\frac{1}{\sqrt{2}}\left(1-\frac{\Omega^{2}/\Delta}{4\Delta_{\rm vec}}+\frac{\Omega^{2}/\Delta}{4V_{\rm int}}-\frac{17(\Omega^{2}/\Delta)^{2}}{32\Delta_{\rm vec}^{2}}\right)\ket{r1}.
\end{eqnarray}
The initial state $\ket{11}$ evolves as 
\begin{eqnarray}
\ket{\Psi} =  e^{\frac{2i\Omega^{2}t}{\Delta}+\frac{i(\Omega^{2}/\Delta)^{2}t}{\Delta_{\rm vec}}\left(1+\frac{2\Omega^{2}/\Delta}{\Delta_{\rm vec}}\right)+\frac{i(\Omega^{2}/\Delta)^{2}t}{2V_{\rm int}}}\times \nonumber \\
\times \left(a_{1}\ket{\Psi_{1}}e^{-i\frac{\Omega^{2}t}{\Delta}\left(1+\frac{\Omega^{2}/\Delta}{8\Delta_{\rm vec}}\right)}+a_{2}\ket{\Psi_{2}}e^{i\frac{\Omega^{2}t}{\Delta}\left(1+\frac{\Omega^{2}/\Delta}{8\Delta_{\rm vec}}\right)}\right),\nonumber 
\end{eqnarray}
where 
\begin{eqnarray}
\label{eq:coefficients_11}
a_{1} & = & -\frac{1}{\sqrt{2}}\left(1-\frac{\Omega^{2}/\Delta}{4\Delta_{\rm vec}}+\frac{\Omega^{2}/\Delta}{4V_{\rm int}}-\frac{17(\Omega^{2}/\Delta)^{2}}{32\Delta_{\rm vec}^{2}}\right),\nonumber \\
a_{2} & = & \frac{1}{\sqrt{2}}\left(1+\frac{\Omega^{2}/\Delta}{4\Delta_{\rm vec}}-\frac{\Omega^{2}/\Delta}{4V_{\rm int}}+\frac{15(\Omega^{2}/\Delta)^{2}}{32\Delta_{\rm vec}^{2}}\right). 
\end{eqnarray} 
After the first $\pi$ pulse the wavefunction, therefore, turns into
\begin{eqnarray}
\ket{\Psi} & = & -e^{\frac{i\pi \Omega^{2}/\Delta}{4\Delta_{\rm vec}}\left(1+\frac{2\Omega^{2}/\Delta}{\Delta_{\rm vec}}\right)+\frac{i\pi \Omega^{2}/\Delta}{4V_{\rm int}}}\times \nonumber \\
&& \times \left(\left(\frac{i\Omega^{2}/\Delta}{2\Delta_{\rm vec}}-\frac{\pi \Omega^{2}/\Delta}{16\Delta_{\rm vec}}-\frac{i\Omega^{2}/\Delta}{2V_{\rm int}}+\frac{i(\Omega^{2}/\Delta)^{2}}{\Delta_{\rm int}^{2}}\right)\ket{11}+\right. \nonumber \\
&& \left. +i\left(1-\left(\frac{1}{8}+\frac{\pi^{2}}{512}\right)\frac{(\Omega^{2}/\Delta)^{2}}{\Delta_{\rm vec}^{2}}\right)\ket{r1}\right). \nonumber
\end{eqnarray}

Next, a $2\pi$ pulse resonant to the target, {\it i.e.} the right atom is applied. The system of equations for the amplitudes $a_{11}$, $a_{1r}$, $a_{r1}$ and $a_{rr}$ is the same 
as Eqs.(\ref{eq:eqs_ind_addr_11_1}) with $\Delta_{\rm vec}\rightarrow -\Delta_{\rm vec}$. The system of equations for $a_{11}$ and $a_{1r}$ is also the same as 
Eqs.(\ref{eq:eqs_ind_addr_11_2}) with $\Delta_{\rm vec}\rightarrow -\Delta_{\rm vec}$. As a result, during the $2\pi$ pulse the state evolves as
\begin{eqnarray}
\ket{\Psi} & = & e^{\frac{2i\Omega^{2}t}{\Delta}-\frac{i(\Omega^{2}/\Delta)^{2}t}{\Delta_{\rm vec}}\left(1-\frac{2\Omega^{2}/\Delta}{\Delta_{\rm vec}}\right)+\frac{i(\Omega^{2}/\Delta)^{2}t}{2V_{\rm int}}}\times \nonumber \\
&& \times \left(a_{1}\ket{\Psi_{1}}e^{-\frac{i\Omega^{2}t}{\Delta}\left(1+\frac{\Omega^{2}/\Delta}{8\Delta_{\rm vec}}\right)}+\right. \nonumber \\
&&\left. +a_{2}\ket{\Psi_{2}}e^{\frac{i\Omega^{2}t}{\Delta}\left(1+\frac{\Omega^{2}/\Delta}{8\Delta_{\rm vec}}\right)}\right),\nonumber 
\end{eqnarray}
where $\ket{\Psi_{1,2}}$ are given by Eqs.(\ref{eq:wavefunct_ind_addr_11}) with $\Delta_{\rm vec}\rightarrow -\Delta_{\rm vec}$ and $\ket{r1}\rightarrow \ket{1r}$,  
$a_{1,2}$ are given by Eqs.(\ref{eq:coefficients_11}). After the $2\pi$ pulse the states $\ket{11}$ and $\ket{r1}$ change as
\begin{eqnarray}
\ket{11}\rightarrow -e^{-\frac{i\pi \Omega^{2}/\Delta}{2\Delta_{\rm vec}}\left(1-\frac{2\Omega^{2}/\Delta}{\Delta_{\rm vec}}\right)+\frac{i\pi \Omega^{2}/\Delta}{2V_{\rm int}}}\times \nonumber \\
\times \left(\ket{11}+\frac{i\pi}{8}\left(\frac{\Omega^{2}/\Delta}{\Delta_{\rm vec}}\right)^{2}\ket{1r}\right), \nonumber \\
\ket{r1}\rightarrow e^{i\pi \frac{\Delta_{\rm vec}}{\Omega^{2}/\Delta}-i\pi \frac{\Omega^{2}/\Delta}{\Delta_{\rm vec}}\left(1+\frac{2\Omega^{2}/\Delta}{\Delta_{\rm vec}}\right)-i\pi \frac{\Omega^{2}/\Delta}{V_{\rm int}}}\ket{r1}. \nonumber 
\end{eqnarray}

As a result, after the $2\pi$ pulse the state becomes
\begin{eqnarray}
\ket{\Psi} & \rightarrow & -e^{\frac{i\pi \Omega^{2}/\Delta}{4\Delta_{\rm vec}}\left(1+\frac{2\Omega^{2}/\Delta}{\Delta_{\rm vec}}\right)+\frac{i\pi \Omega^{2}/\Delta}{4V_{\rm int}}}\times \nonumber \\
&& \times \left[-\left(\left(\frac{i}{2}-\frac{\pi}{16}\right)\frac{\Omega^{2}/\Delta}{\Delta_{\rm vec}}-\frac{i\Omega^{2}/\Delta}{2V_{\rm int}}+i\frac{(\Omega^{2}/\Delta)^{2}}{\Delta_{\rm vec}^{2}}\right)\times \right. \nonumber \\
&&\left. \times e^{-\frac{i\pi \Omega^{2}/\Delta}{2\Delta_{\rm vec}}\left(1-\frac{2\Omega^{2}/\Delta}{\Delta_{\rm vec}}\right)+\frac{i\pi \Omega^{2}/\Delta}{2V_{\rm int}}}\ket{11}+ \right. \nonumber \\
&& \left. +i\left(1-\left(\frac{1}{8}+\frac{\pi^{2}}{512}\right)\frac{(\Omega^{2}/\Delta)^{2}}{\Delta_{\rm vec}^{2}}\right)\times \right. \nonumber \\
&&\left. \times e^{i\pi \frac{\Delta_{\rm vec}}{\Omega^{2}/\Delta}-\frac{i\pi \Omega^{2}/\Delta}{\Delta_{\rm vec}}\left(1+\frac{2\Omega^{2}/\Delta}{\Delta_{\rm vec}}\right)-\frac{i\pi \Omega^{2}/\Delta}{V_{\rm int}}}\ket{r1}\right], \nonumber
\end{eqnarray}
where we neglected the small $\frac{(\Omega^{2}/\Delta)^{3}}{\Delta_{\rm vec}^{3}}\ket{1r}$ term.

Finally, after the second $\pi$ pulse resonant to the left atom, the states $\ket{11}$ and $\ket{r1}$ evolve as 
\begin{eqnarray}
\ket{11} & \rightarrow & a_{1}\ket{\Psi_{1}}e^{-i\lambda_{1}t}+a_{2}\ket{\Psi_{2}}e^{-i\lambda_{2}t}, \nonumber \\
\ket{r1} & \rightarrow & a_{2}\ket{\Psi_{1}}e^{-i\lambda_{1}t}-a_{1}\ket{\Psi_{2}}e^{-i\lambda_{2}t}. \nonumber
\end{eqnarray} 

At the end of the phase gate the initial $\ket{11}$ state evolves into
\begin{eqnarray}
\ket{\Psi} & \rightarrow & -e^{i\frac{\pi \Omega^{2}/\Delta}{\Delta_{\rm vec}}\left(1+\frac{2\Omega^{2}/\Delta}{\Delta_{\rm vec}}\right)+i\frac{\pi \Omega^{2}/\Delta}{V_{\rm int}}}\times \nonumber \\
&& \times \left[\left(1-i\pi\frac{\Omega^{2}/\Delta}{\Delta_{\rm vec}}-i\pi \frac{\Omega^{2}/\Delta}{V_{\rm int}} \right. \right.  \nonumber \\
&&  \left. -\left(\frac{33i\pi}{16}+\frac{\pi^{2}}{2}\right)\frac{(\Omega^{2}/\Delta)^{2}}{\Delta_{\rm vec}^{2}}\right)\ket{11}  \nonumber \\
&&  -\left(\frac{\Omega^{2}/\Delta}{\Delta_{\rm vec}}-\frac{\Omega^{2}/\Delta}{V_{\rm int}}+ \right. \nonumber \\
&&\left. \left. +\left(\frac{\pi^{2}}{32}+\frac{3\pi i}{4}-2\right)\frac{(\Omega^{2}/\Delta)^{2}}{\Delta_{\rm vec}^{2}}\right)\ket{r1}\right]. \nonumber
\end{eqnarray}

The projection on the desired $-\ket{11}$ state is then
\begin{eqnarray}
-\langle 11 \vert \Psi\rangle & = & e^{i\frac{\pi \Omega^{2}/\Delta}{\Delta_{\rm vec}}\left(1+\frac{2\Omega^{2}/\Delta}{\Delta_{\rm vec}}\right)+i\frac{\pi \Omega^{2}/\Delta}{V_{\rm int}}} \nonumber \\
&& \left(1-i\pi\frac{\Omega^{2}/\Delta}{\Delta_{\rm vec}}-i\pi \frac{\Omega^{2}/\Delta}{V_{\rm int}}\right. \nonumber \\
&&\left. -\left(\frac{33\pi i}{16}+\frac{\pi^{2}}{2}\right)\frac{(\Omega^{2}/\Delta)^{2}}{\Delta_{\rm vec}^{2}}\right). \nonumber
\end{eqnarray}
The error can be found from $\left|-\langle 11 \vert \Psi \rangle\right|^{2}\sim 1+O\left((\Omega^{2}/\Delta)^{2}/\Delta_{\rm vec}V_{\rm int}\right)$.

\subsubsection{$\ket{01}$ initial state}

Now we analyze the evolution of the $\ket{01}$ state. During the first $\pi$ pulse, resonant to the left atom, the amplitudes of the states $\ket{01}$ 
and $\ket{0r}$ (other amplitudes are much smaller) evolve according to the Hamiltonian
\begin{equation}
H/\hbar=
\left(\begin{array}{cc}
-\frac{2\Omega_{1}^{2}}{\Delta+\Delta_{hf}} & -\frac{\Omega_{1}\Omega_{2}}{\Delta+\Delta_{hf}}\\
-\frac{\Omega_{1}\Omega_{2}}{\Delta+\Delta_{hf}} & \left(\Delta_{\rm vec}-\frac{\Omega_{1}^{2}}{\Delta+\Delta_{hf}+\Delta_{\rm vec}}-\frac{\Omega_{2}^{2}}{\Delta+\Delta_{\rm vec}}\right)
\end{array} \right)
\end{equation}
The initial state $\ket{01}$ changes as 
\begin{eqnarray}
\ket{\Psi} & = & e^{-i\Delta_{\rm vec}t+i\frac{\Omega^{2}t}{\Delta}+i\frac{\Omega^{2}t}{\Delta+\Delta_{hf}}}\times \nonumber \\
&& \times \left(-\frac{\Omega^{2}/\Delta}{\Delta_{\rm vec}}\ket{\Psi_{+}}e^{-i\frac{\Delta_{\rm vec}t}{2}\left(1+2\frac{(\Omega^{2}/\Delta)^{2}}{\Delta_{\rm vec}^{2}}\right)}+\right. \nonumber \\
&&\left. +\left(1-\frac{1}{2}\frac{(\Omega^{2}/\Delta)^{2}}{\Delta_{\rm vec}^{2}}\right)\ket{\Psi_{-}}e^{i\frac{\Delta_{\rm vec}t}{2}\left(1+2\frac{(\Omega^{2}/\Delta)^{2}}{\Delta_{\rm vec}^{2}}\right)}\right),\nonumber 
\end{eqnarray}
where the eigenstates are given by
\begin{eqnarray}
\label{eq:Psipm}
\ket{\Psi_{+}}& = & -\frac{\Omega^{2}/\Delta}{\Delta_{\rm vec}}\ket{01}+\left(1-\frac{(\Omega^{2}/\Delta)^{2}}{\Delta_{\rm vec}^{2}}\right)\ket{0r},\nonumber \\
\ket{\Psi_{-}}& = & \left(1-\frac{(\Omega^{2}/\Delta)^{2}}{\Delta_{\rm vec}^{2}}\right)\ket{01}+\frac{\Omega^{2}/\Delta}{\Delta_{\rm vec}}\ket{0r}.\nonumber \\
\end{eqnarray}

After the first $\pi$ pulse the state turns into 
\begin{eqnarray*}
\ket{\Psi}\rightarrow 
ie^{i\frac{\pi \Delta}{2(\Delta+\Delta_{hf})}}\left[\left(1+i\frac{\pi \Omega^{2}/\Delta}{\Delta_{\rm vec}}+\right. \right. \nonumber \\
\left. \left.+(e^{-i\pi \Delta_{\rm vec}/(\Omega^{2}/\Delta)}-1-\frac{\pi^{2}}{8})\frac{(\Omega^{2}/\Delta)^{2}}{\Delta_{\rm vec}^{2}}\right)\ket{01}+ \right. \nonumber \\
\left. +\frac{\Omega^{2}/\Delta}{\Delta_{\rm vec}}\left(1+i\frac{\pi \Omega^{2}/\Delta}{2\Delta_{\rm vec}}-\left(1-\frac{i\pi \Omega^{2}/\Delta}{\Delta_{\rm vec}}\right)e^{-i\frac{\pi \Delta_{\rm vec}}{\Omega^{2}/\Delta}}\right)\ket{0r}\right]. \nonumber
\end{eqnarray*}

During the $2\pi$ pulse resonant to the right atom, the $\ket{01}$ and $\ket{0r}$ states evolve as
\begin{eqnarray}
\ket{01} & \rightarrow & e^{i\frac{\Omega^{2}t}{\Delta+\Delta_{hf}}+i\frac{\Omega^{2}t}{\Delta}}\frac{1}{2}\left((\ket{01}-\ket{0r})e^{-i\Omega^{2}t/\Delta}+\right. \nonumber \\
&&\left. +(\ket{01}+\ket{0r})e^{i\Omega^{2}t/\Delta}\right), \nonumber \\
\ket{0r} & \rightarrow & e^{i\frac{\Omega^{2}t}{\Delta+\Delta_{hf}}+i\frac{\Omega^{2}t}{\Delta}}\frac{1}{2}\left((\ket{01}+\ket{0r})e^{i\Omega^{2}t/\Delta}\right. \nonumber \\
&&\left.-(\ket{01}-\ket{0r})e^{-i\Omega^{2}t/\Delta}\right). \nonumber
\end{eqnarray}

As a result, after the $2\pi$ pulse 
\begin{eqnarray}
\ket{01} \rightarrow \ket{01}e^{i\pi \Delta/(\Delta+\Delta_{hf})},\nonumber \\
\ket{0r} \rightarrow \ket{0r}e^{i\pi \Delta/(\Delta+\Delta_{hf})}, \nonumber
\end{eqnarray}

During the second $\pi$ pulse the states $\ket{01}$ and $\ket{0r}$ change in the following way
\begin{eqnarray}
\ket{01} & \rightarrow & ia_{+}\ket{\Psi_{+}}e^{-i\frac{\pi \Delta_{\rm vec}}{2\Omega^{2}/\Delta}+i\frac{\pi \Delta}{\Delta+\Delta_{hf}}-i\frac{\pi \Omega^{2}/\Delta}{\Delta_{\rm vec}}}+\nonumber \\
&& +ia_{-}\ket{\Psi_{-}}e^{i\frac{\pi \Delta}{\Delta+\Delta_{hf}}+i\frac{\pi \Omega^{2}/\Delta}{\Delta_{\rm vec}}}, \nonumber \\
\ket{0r} & \rightarrow & ia_{-}\ket{\Psi_{+}}e^{-i\frac{\pi \Delta_{\rm vec}}{2\Omega^{2}/\Delta}+i\frac{\pi \Delta}{\Delta+\Delta_{hf}}-i\frac{\pi \Omega^{2}/\Delta}{\Delta_{\rm vec}}} \nonumber \\
&& -a_{+}\ket{\Psi_{-}}e^{i\frac{\pi \Delta}{\Delta+\Delta_{hf}}+i\frac{\pi \Omega^{2}/\Delta}{\Delta_{\rm vec}}}, \nonumber
\end{eqnarray}
where $\ket{\Psi_{\pm}}$ are given by Eqs.(\ref{eq:Psipm}) and $a_{+}=-(\Omega^{2}/\Delta)/\Delta_{\rm vec}$, $a_{-}=1-\frac{(\Omega^{2}/\Delta)^{2}}{2\Delta_{\rm vec}^{2}}$.
After the second $\pi$ pulse the initial $\ket{01}$ state becomes
\begin{eqnarray}
\ket{\Psi} & \rightarrow & -e^{-2\pi i \Delta_{hf}/\Delta}\left[\left(1+i\pi \frac{\Omega^{2}/\Delta}{\Delta_{\rm vec}}+\right. \right. \nonumber \\
&& \left.  +\left(e^{-i\pi \frac{\Delta_{\rm vec}}{\Omega^{2}/\Delta}}-2-\frac{\pi^{2}}{2}\right)\frac{(\Omega^{2}/\Delta)^{2}}{\Delta_{\rm vec}^{2}}\right)\ket{01}+ \nonumber \\
&& +\left(\frac{\Omega^{2}/\Delta}{\Delta_{\rm vec}}\left(1-e^{-i\frac{\pi \Delta_{\rm vec}}{\Omega^{2}/\Delta}}\right)+ \right. \nonumber \\
&& \left. \left.+i\pi \frac{(\Omega^{2}/\Delta)^{2}}{\Delta_{\rm vec}^{2}}\left(1+e^{-i\pi\frac{\Delta_{\rm vec}}{\Omega^{2}/\Delta}}\right)\right)\ket{0r}\right]. \nonumber 
\end{eqnarray}
The projection on the $-\ket{01}$ state is then
\begin{eqnarray}
-\langle 01 \vert \Psi\rangle & = & e^{-2\pi i \Delta_{hf}/\Delta}\left(1+i\pi \frac{\Omega^{2}/\Delta}{\Delta_{\rm vec}}+\right. \nonumber \\ 
&&\left. +\left(e^{-i\frac{\pi \Delta_{\rm vec}}{\Omega^{2}/\Delta}}-2-\frac{\pi^{2}}{2}\right)\frac{(\Omega^{2}/\Delta)^{2}}{\Delta_{\rm vec}^{2}}\right). \nonumber
\end{eqnarray}
Assuming $\pi \Delta_{\rm vec}/(\Omega^{2}/\Delta)=2\pi n$ we have $\left|-\langle 01 \vert \Psi \rangle\right|^{2}=1-2\frac{(\Omega^{2}/\Delta)^{2}}{\Delta_{\rm vec}^{2}}$, 
and the resulting error is $2\frac{(\Omega^{2}/\Delta)^{2}}{\Delta_{\rm vec}^{2}}$.

\subsubsection{$\ket{10}$ initial state}
During the first $\pi$ pulse the amplitudes $a_{10}$ and $a_{r0}$ (other states are far detuned and their amplitudes are much smaller) 
change according to the Hamiltonian
\begin{equation}
H/\hbar=
\left(\begin{array}{cc}
-\left(\frac{\Omega_{1}^{2}}{\Delta}+\frac{\Omega_{1}^{2}}{\Delta+\Delta_{hf}}\right) & -\frac{\Omega_{1}\Omega_{2}}{\Delta}\\
-\frac{\Omega_{1}\Omega_{2}}{\Delta} & -\left(\frac{\Omega_{2}^{2}}{\Delta}+\frac{\Omega_{1}^{2}}{\Delta+\Delta_{hf}}\right)
\end{array} \right)
\end{equation}
During the first $\pi$ pulse the states $\ket{01}$ and $\ket{0r}$ evolve as 
\begin{eqnarray}
\ket{10} & \rightarrow & e^{i\frac{\Omega^{2}t}{\Delta+\Delta_{hf}}+i\frac{\Omega^{2}t}{\Delta}}\frac{1}{2}\left((\ket{10}-\ket{r0})e^{-i\frac{\Omega^{2}t}{\Delta}}+ \right. \nonumber \\
&& \left. +(\ket{10}+\ket{r0})e^{i\frac{\Omega^{2}t}{\Delta}}\right), \nonumber \\
\ket{r0} & \rightarrow & e^{i\frac{\Omega^{2}t}{\Delta+\Delta_{hf}}+i\frac{\Omega^{2}t}{\Delta}}\frac{1}{2}\left((\ket{10}+\ket{r0})e^{i\frac{\Omega^{2}t}{\Delta}} \right. \nonumber \\
&& \left.-(\ket{10}-\ket{r0})e^{-i\frac{\Omega^{2}t}{\Delta}}\right).  \nonumber
\end{eqnarray}
After the first $\pi$ pulse the states $\ket{10}$, $\ket{r0}$ become 
\begin{eqnarray}
\ket{10} & \rightarrow & -ie^{-i\frac{\pi \Delta_{hf}}{2\Delta}}\ket{r0}, \nonumber \\
\ket{r0} & \rightarrow & -ie^{-i\frac{\pi \Delta_{hf}}{2\Delta}}\ket{10}. \nonumber
\end{eqnarray} 

During the $2\pi$ pulse resonant to the right atom, the amplitudes $a_{10}$ and $a_{r0}$ are governed by the Hamiltonian 
\begin{widetext}
\begin{equation}
H/\hbar=
\left(\begin{array}{cc}
-\left(\frac{\Omega_{1}^{2}}{\Delta-\Delta_{\rm vec}}+\frac{\Omega_{1}^{2}}{\Delta+\Delta_{hf}}\right) & -\frac{\Omega_{1}\Omega_{2}}{\Delta-\Delta_{\rm vec}}\\
-\frac{\Omega_{1}\Omega_{2}}{\Delta-\Delta_{\rm vec}} & -\left(\Delta_{\rm vec}+\frac{\Omega_{1}^{2}}{\Delta+\Delta_{hf}-\Delta_{\rm vec}}+\frac{\Omega_{2}^{2}}{\Delta-\Delta_{\rm vec}}\right)
\end{array} \right)
\end{equation}
\end{widetext}
As a result, the state $\ket{r0}$ evolves as
\begin{eqnarray}
\ket{r0} & \rightarrow & e^{i\frac{\Omega^{2}t}{\Delta}+i\frac{\Omega^{2}}{\Delta+\Delta_{hf}}}\left(-\frac{\Omega^{2}/\Delta}{\Delta_{\rm vec}}\ket{\Psi^{+}}e^{-i\frac{(\Omega^{2}/\Delta)^{2}t}{\Delta_{\rm vec}}}+\right. \nonumber \\
&& \left. +\left(1-\frac{(\Omega^{2}/\Delta)^{2}}{2\Delta_{\rm vec}^{2}}\right)\ket{\Psi^{-}}e^{i\Delta_{\rm vec}t+i\frac{(\Omega^{2}/\Delta)^{2}t}{\Delta_{\rm vec}}}\right), \nonumber
\end{eqnarray}
where
\begin{eqnarray}
\ket{\Psi^{+}} & = & \left(1-\frac{(\Omega^{2}/\Delta)^{2}}{\Delta_{\rm vec}^{2}}\right)\ket{10}-\frac{\Omega^{2}/\Delta}{\Delta_{\rm vec}}\ket{r0}, \nonumber \\
\ket{\Psi^{-}} & = & \frac{\Omega^{2}/\Delta}{\Delta_{\rm vec}}\ket{10}+\left(1-\frac{(\Omega^{2}/\Delta)^{2}}{\Delta_{\rm vec}^{2}}\right)\ket{r0}. \nonumber
\end{eqnarray}

After the second $\pi$ pulse the initial state $\ket{10}$ evolves into
\begin{eqnarray*}
\ket{\Psi}\rightarrow -e^{-2\pi i\frac{\Delta_{hf}}{\Delta}}\times \nonumber \\
\times \left[\left(1+i\pi\frac{\Omega^{2}/\Delta}{\Delta_{\rm vec}}\right)e^{i\pi \frac{\Delta_{\rm vec}}{\Omega^{2}/\Delta}}+\right. \nonumber \\
\left. +\left(1-e^{i\pi \frac{\Delta_{\rm vec}}{\Omega^{2}/\Delta}}\left(1+\frac{\pi^{2}}{2}\right)\right)\frac{(\Omega^{2}/\Delta)^{2}}{\Delta_{\rm vec}^{2}}\right)\ket{10}+ \nonumber \\
 +\left(e^{i\pi \frac{\Delta_{\rm vec}}{\Omega^{2}/\Delta}}\frac{\Omega^{2}/\Delta}{\Delta_{\rm vec}}\left(1+i\pi \frac{\Omega^{2}/\Delta}{\Delta_{\rm vec}}\right)\right. \nonumber \\
\left. \left. -\frac{\Omega^{2}/\Delta}{\Delta_{\rm vec}}\left(1-i\pi \frac{\Omega^{2}/\Delta}{\Delta_{\rm vec}}\right)\right)\ket{r0}\right]. \nonumber
\end{eqnarray*}

The overlap with the desired $-\ket{10}$ state is then
\begin{eqnarray} 
-\langle 10 \vert \Psi\rangle=e^{-2\pi i\frac{\Delta_{hf}}{\Delta}}\left(\frac{(\Omega^{2}/\Delta)^{2}}{\Delta_{\rm vec}^{2}}+\right. \nonumber \\
\left. +e^{i\pi\frac{\Delta_{\rm vec}}{\Omega^{2}/\Delta}}\left(1+i\pi \frac{\Omega^{2}/\Delta}{\Delta_{\rm vec}}-\left(1+\frac{\pi^{2}}{2}\right)\frac{(\Omega^{2}/\Delta)^{2}}{\Delta_{\rm vec}^{2}}\right)\right). \nonumber
\end{eqnarray}
If $\pi \Delta_{\rm vec}/(\Omega^{2}/\Delta)=2\pi n$
\begin{eqnarray}
-\langle 10 \vert \Psi\rangle=e^{-2\pi i \Delta_{hf}/\Delta}\left(1+i\pi \frac{\Omega^{2}/\Delta}{\Delta_{\rm vec}}-\frac{\pi^{2}}{2}\frac{(\Omega^{2}/\Delta)^{2}}{\Delta_{\rm vec}^{2}}\right), \nonumber
\end{eqnarray}
resulting in $\left|-\langle 10 \vert \Psi\rangle \right|^{2}=1+O\left(\left(\frac{\Omega^{2}/\Delta}{\Delta_{\rm vec}}\right)^{4}\right)$.

Finally, the $\ket{00}$ state evolves into $\ket{00}\rightarrow \ket{00}e^{2i\pi\Delta/(\Delta +\Delta_{hf})}$ after the gate and $\left|\langle 00 \vert \Psi\rangle\right|^{2}=1$.

The error averaged over all initial states is then $\epsilon=\frac{(\Omega^{2}/\Delta)^{2}}{2\Delta_{\rm vec}^{2}}$.

\subsection{Error due to the decay of Rydberg states}

We again start the analysis from the initial state $\ket{11}$. After the first $\pi$ pulse resonant to the left atom it becomes 
\begin{eqnarray}
\ket{\Psi} \rightarrow -ie^{-\gamma \pi/(2\Omega^{2}/\Delta)}\ket{r1}. \nonumber
\end{eqnarray}

At the end of the $2\pi$ pulse resonant to the right atom the state $\ket{r1}$ is given by
\begin{eqnarray}
\ket{r1} \rightarrow e^{i\pi \Delta_{\rm vec}/(\Omega^{2}/\Delta)-\gamma \pi/(\Omega^{2}/\Delta)}\ket{r1}. \nonumber
\end{eqnarray}
Finally, after the second $\pi$ pulse resonant to the left atom,
\begin{eqnarray}
\ket{r1} \rightarrow -ie^{-\pi \gamma /(2\Omega^{2}/\Delta)}\ket{11}. \nonumber
\end{eqnarray}
As a result, at the end of the gate 
\begin{eqnarray}
\ket{\Psi} \rightarrow -e^{i\pi \Delta_{\rm vec}/(\Omega^{2}/\Delta)-2\pi \gamma/(\Omega^{2}/\Delta)}\ket{11}. \nonumber
\end{eqnarray}
From the overlap $\left|-\langle 11 \vert \Psi\rangle \right|^{2}=e^{-4\pi \gamma/(\Omega^{2}/\Delta)}$ the error is $4\pi \gamma/(\Omega^{2}/\Delta)$. 

The initial state $\ket{\Psi}=\ket{01}$ after the first $\pi$ pulse turns into $\ket{\Psi}\rightarrow -\ket{01}$. After the $2\pi$ pulse $\ket{\Psi}\rightarrow e^{-i\pi \Delta_{hf}/\Delta}\ket{01}$. 
Finally, after the second $\pi$ pulse $\ket{\Psi} \rightarrow -e^{-i\pi \Delta_{hf}/\Delta}\ket{01}$, giving $\left|-\langle 01 \vert \Psi\rangle \right|^{2}=1$.

The initial state $\ket{\Psi}=\ket{10}$ after the first $\pi$ pulse evolves into $\ket{\Psi} \rightarrow -ie^{-\gamma \pi/(2\Omega^{2}/\Delta)}\ket{r0}$. 
After the $2\pi$ pulse $\ket{r0} \rightarrow e^{-i\pi \Delta_{hf}/\Delta+i\pi \Delta_{\rm vec}/(\Omega^{2}/\Delta)-\gamma \pi/(\Omega^{2}/\Delta)}\ket{r0}$. 
The state $\ket{r0}$ turns into $\ket{r0} \rightarrow -ie^{-i\pi \Delta_{hf}/\Delta}\ket{10}$ after the second $\pi$ pulse. As a result, at the of the 
gate $\ket{\Psi} \rightarrow -e^{-2\pi i \Delta_{hf}/\Delta+i\pi \Delta_{\rm vec}/(\Omega^{2}/\Delta)-3\pi \gamma/(2\Omega^{2}/\Delta)}\ket{10}$. This 
gives $\left|-\langle 10 \vert \Psi\rangle \right|^{2}=e^{-3\pi \gamma/(\Omega^{2}/\Delta)}$ resulting in the error $3\pi \gamma/(\Omega^{2}/\Delta)$. Finally, 
the initial state $\ket{00}\rightarrow \ket{00}$ after the gate and there is no error due to the Rydberg state decay.

The error averaged over all initial qubit states is then $7\pi \gamma/(4\Omega^{2}/\Delta)$.

\subsection{Error due to the variation of the Rabi frequency}

\subsubsection{$\ket{11}$ initial state}

First, we analyze the $\ket{11}$ initial state. During the first $\pi$ pulse, resonant to the left atom, the state evolves as
\begin{eqnarray}
\ket{\Psi}=e^{2i\Omega^{2}t/\Delta}\left(\frac{\ket{11}-\ket{r1}}{2}e^{-i\Omega^{2}t/\Delta}+\frac{\ket{11}+\ket{r1}}{2}e^{i\Omega^{2}t/\Delta}\right). \nonumber
\end{eqnarray}
Again, we assume that $\Omega=\langle \Omega \rangle +\delta \Omega$ and $\langle \Omega \rangle ^{2}T/\Delta=\pi/2$ during a $\pi$ pulse. As a result, 
after the first $\pi$ pulse 
\begin{eqnarray}
\ket{\Psi} \rightarrow  -\frac{i}{2}e^{2\pi i\delta \Omega/\langle \Omega \rangle+i\pi (\delta \Omega)^{2}/\langle \Omega \rangle^{2}}\times \nonumber \\
\left(\begin{array}{cc}
2\pi i \frac{\delta \Omega}{\langle \Omega \rangle}+i\pi \frac{(\delta \Omega)^{2}}{\langle \Omega \rangle ^{2}} & 0 \\
0 & 2-\pi^{2}\frac{(\delta \Omega)^{2}}{\langle \Omega \rangle ^{2}}
\end{array} \right)
\left(\begin{array}{c} \ket{11} \\ \ket{r1}
\end{array} \right) \nonumber
\end{eqnarray}
During the $2\pi$ pulse resonant to the right atom, the states $\ket{11}$ and $\ket{r1}$ evolve as
\begin{eqnarray}
\ket{11} & \rightarrow & e^{2i\Omega^{2}t/\Delta}\left(\frac{\ket{11}-\ket{1r}}{2}e^{-2i\Omega^{2}t/\Delta}+\right. \nonumber \\
\left. +\frac{\ket{11}+\ket{1r}}{2}e^{2i\Omega^{2}t/\Delta}\right), \nonumber \\
\ket{r1} & \rightarrow & \ket{r1}e^{i \Delta_{\rm vec}t-2\pi i \Omega^{2}t/\Delta}, \nonumber 
\end{eqnarray}
so that at the end of the $2\pi$ pulse
\begin{eqnarray}
\ket{11}\rightarrow -\frac{1}{2}e^{4\pi i\delta \Omega/\langle \Omega \rangle +2\pi i (\delta \Omega)^{2}/\langle \Omega \rangle ^{2}}\times \nonumber \\
\left(\begin{array}{cc}
2-4\pi^{2}\frac{(\delta \Omega)^{2}}{\langle \Omega \rangle^{2}} & 0 \\
0 & 4\pi i \frac{\delta \Omega}{\langle \Omega \rangle}+2\pi i\frac{(\delta \Omega)^{2}}{\langle \Omega \rangle ^{2}}
\end{array} \right)
\left(\begin{array}{c} \ket{11} \\ \ket{1r}
\end{array} \right) \nonumber
\end{eqnarray}
The initial state $\ket{11}$ after the $2\pi$ pulse becomes
\begin{widetext}
\begin{eqnarray}
\ket{\Psi} \rightarrow -\frac{1}{2}e^{6\pi i \delta \Omega/\langle \Omega \rangle +3i\pi (\delta \Omega)^{2}/\langle \Omega \rangle ^{2}} \times  \nonumber \\
\left(\begin{array}{ccc}
-\left(2\pi i \frac{\delta \Omega}{\langle \Omega \rangle}+\pi i\frac{(\delta \Omega)^{2}}{\langle \Omega \rangle^{2}}\right) & 0 & 0 \\
0 & 4\pi^{2}\frac{(\delta \Omega)^{2}}{\langle \Omega \rangle ^{2}} & 0 \\
0 & 0 & e^{i\pi \Delta_{\rm vec}/(\Omega^{2}/\Delta)-4\pi i \delta \Omega/\langle \Omega \rangle -2\pi i (\delta \Omega)^{2}/\langle \Omega \rangle^{2}}\left(2-8\pi i\frac{\delta \Omega}{\langle \Omega \rangle}-(4\pi i+17\pi^{2}) \frac{(\delta \Omega)^{2}}{\langle \Omega \rangle^{2}}\right)
\end{array} \right)
\left(\begin{array}{c} \ket{11} \\ \ket{1r} \\ \ket{r1}
\end{array} \right) \nonumber
\end{eqnarray}
\end{widetext}
After the second $\pi$ pulse resonant to the left atom the states $\ket{11}$, $\ket{r1}$ and $\ket{1r}$ become
\begin{widetext}
\begin{eqnarray}
\left(\begin{array}{c} \ket{11} \\ \ket{r1}
\end{array}\right)
\rightarrow
-\frac{i}{2}\left(\begin{array}{cc}
2\pi i \frac{\delta \Omega}{\langle \Omega \rangle} +i\pi \frac{(\delta \Omega)^{2}}{\langle \Omega \rangle^{2}} & 2-\pi^{2}\frac{(\delta \Omega)^{2}}{\langle \Omega \rangle^{2}} \\
2-\pi^{2}\frac{(\delta \Omega)^{2}}{\langle \Omega \rangle^{2}} & 2\pi i \frac{\delta \Omega}{\langle \Omega \rangle}+i\pi \frac{(\delta \Omega)^{2}}{\langle \Omega \rangle^{2}}
\end{array} \right)
\left(\begin{array}{c} \ket{11} \\ \ket{r1}
\end{array} \right) \nonumber \\
\ket{1r} \rightarrow \ket{1r}e^{-i\pi \Delta_{\rm vec}/(\Omega^{2}/\Delta)-4\pi i \delta \Omega/\langle \Omega \rangle -2\pi i (\delta \Omega)^{2}/\langle \Omega \rangle^{2}}. \nonumber 
\end{eqnarray}
\end{widetext}
If $\pi \Delta_{\rm vec}/(\Omega^{2}/\Delta)=2\pi n$ the wavefunction after the $2\pi$ pulse is given by
\begin{widetext}
\begin{eqnarray}
\ket{\Psi}\rightarrow -\frac{i}{2}e^{2\pi i \delta \Omega/\langle \Omega \rangle +i\pi (\delta \Omega)^{2}/\langle \Omega \rangle^{2}}\times \nonumber \\
\left(\begin{array}{ccc}
-\frac{i}{2}\left(4-16\pi i\frac{\delta \Omega}{\langle \Omega \rangle} -(8\pi i +32\pi^{2})\frac{(\delta \Omega)^{2}}{\langle \Omega \rangle^{2}}\right) & 0 & 0 \\
0 & 4\pi^{2}\frac{(\delta \Omega)^{2}}{\langle \Omega \rangle^{2}} & 0 \\
0 & 0 & -16\pi^{2}i\frac{(\delta \Omega)^{2}}{\langle \Omega \rangle^{2}}
\end{array} \right)
\left(\begin{array}{c} \ket{11} \\ \ket{1r} \\ \ket{r1}
\end{array} \right)
\end{eqnarray}
\end{widetext}
The projection to the desired $-\ket{11}$ state is then
\begin{eqnarray}
-\langle 11 \vert \Psi\rangle =e^{2\pi i \delta \Omega/\langle \Omega \rangle +i\pi (\delta \Omega)^{2}/\langle \Omega \rangle^{2}}\times \nonumber \\
\times \left(1-4\pi i\frac{\delta \Omega}{\langle \Omega \rangle}-(2\pi i+8\pi^{2})\frac{(\delta \Omega)^{2}}{\langle \Omega \rangle^{2}}\right), \nonumber  
\end{eqnarray}
giving $\left|-\langle 11 \vert \Psi \rangle \right|^{2}=1+O\left(\left(\frac{\delta \Omega}{\langle \Omega \rangle}\right)^{4}\right)$.

\subsubsection{$\ket{01}$ initial state}

Next, the initial $\ket{01}$ state during the first $\pi$ pulse evolves as
\begin{eqnarray}
\ket{\Psi}=e^{2i\frac{\Omega^{2}t}{\Delta}-i\frac{\Omega^{2}t}{\Delta}\frac{\Delta_{hf}}{\Delta}}\ket{01}. \nonumber 
\end{eqnarray}
After the $\pi$ pulse
\begin{eqnarray}
\ket{\Psi} \rightarrow -e^{2\pi i\frac{\delta \Omega}{\langle \Omega \rangle}+i\pi \frac{(\delta \Omega)^{2}}{\langle \Omega \rangle^{2}}-\frac{i\pi \Delta_{hf}}{2\Delta}}\ket{01}. \nonumber
\end{eqnarray} 

During the $2\pi$ pulse resonant to the right atom, the $\ket{01}$ state changes as
\begin{eqnarray}
\ket{01} \rightarrow e^{2i\frac{\Omega^{2}t}{\Delta}-i\frac{\Omega^{2}t}{\Delta}\frac{\Delta_{hf}}{\Delta}}\frac{1}{2}\left((\ket{01}-\ket{0r})e^{-i\Omega^{2}t/\Delta}+ \right. \nonumber \\
\left. +(\ket{01}+\ket{0r})e^{i\Omega^{2}t/\Delta}\right). \nonumber
\end{eqnarray}
After the $2\pi$ pulse the initial state becomes
\begin{eqnarray}
\ket{\Psi}\rightarrow e^{6\pi i\frac{\delta \Omega}{\langle \Omega \rangle}+3\pi i \frac{(\delta \Omega)^{2}}{\langle \Omega \rangle^{2}}-\frac{3\pi i \Delta_{hf}}{2\Delta}}\times \nonumber \\
\left(\begin{array}{cc}
1-2\pi^{2}\frac{(\delta \Omega)^{2}}{\langle \Omega \rangle^{2}} & 0 \\
0 & 2\pi i \frac{\delta \Omega}{\langle \Omega \rangle}+i\pi \frac{(\delta \Omega)^{2}}{\langle \Omega \rangle^{2}}
\end{array} \right)
\left(\begin{array}{c} \ket{01} \\ \ket{0r}
\end{array} \right)
\end{eqnarray}
During the second $\pi$ pulse resonant to the left atom, the states $\ket{01}$ and $\ket{0r}$ evolve as
\begin{eqnarray}
\ket{01} & \rightarrow & e^{2i\frac{\Omega^{2}t}{\Delta}-\frac{i\Omega^{2}t}{\Delta}\frac{\Delta_{hf}}{\Delta}}\ket{01}, \nonumber \\
\ket{0r} & \rightarrow & e^{-i\Delta_{\rm vec}t+2i\frac{\Omega^{2}t}{\Delta}-i\frac{\Omega^{2}t}{\Delta}\frac{\Delta_{hf}}{\Delta}}\ket{0r}. \nonumber
\end{eqnarray}
As a result, at the end of the gate the initial state $\ket{01}$ turns into
\begin{eqnarray}
\ket{\Psi} \rightarrow -e^{8\pi i\frac{\delta \Omega}{\langle \Omega \rangle}+4\pi i \frac{(\delta \Omega)^{2}}{\langle \Omega \rangle^{2}}-2\pi i\frac{\Delta_{hf}}{\Delta}}\times \nonumber \\
\left(\begin{array}{cc}
1-2\pi^{2}\frac{(\delta \Omega)^{2}}{\langle \Omega \rangle^{2}} & 0 \\
0 & 2\pi i \frac{\delta \Omega}{\langle \Omega \rangle}+i\pi \frac{(\delta \Omega)^{2}}{\langle \Omega \rangle^{2}}
\end{array} \right)
\left(\begin{array}{c} \ket{01} \\ \ket{0r}
\end{array} \right) \nonumber
\end{eqnarray}
From $\left|-\langle 01 \vert \Psi\rangle \right|^{2}=1-4\pi^{2}\frac{(\delta \Omega)^{2}}{\langle \Omega \rangle^{2}}$ the error is $4\pi^{2}(\delta \Omega)^{2}/\langle \Omega \rangle^{2}$.

\subsubsection{$\ket{10}$ initial state}

The initial state $\ket{10}$ during the first $\pi$ pulse evolves as
\begin{eqnarray}
\ket{\Psi}=e^{2i \frac{\Omega^{2}t}{\Delta}-i\frac{\Omega^{2}t}{\Delta}\frac{\Delta_{hf}}{\Delta}}\frac{1}{2}\left(\left(\ket{10}-\ket{r0}\right)e^{-i\Omega^{2}t/\Delta}+\right. \nonumber \\
\left.+\left(\ket{10}+\ket{r0}\right)e^{i\Omega^{2}t/\Delta}\right). \nonumber
\end{eqnarray}
After the $\pi$ pulse 
\begin{eqnarray}
\ket{\Psi} \rightarrow -\frac{i}{2}e^{2\pi i\frac{\delta \Omega}{\langle \Omega \rangle}+i\pi \frac{(\delta \Omega)^{2}}{\langle \Omega \rangle^{2}}-\frac{i\pi \Delta_{hf}}{\Delta}}\times \nonumber \\
\left(\begin{array}{cc}
2\pi i \frac{\delta \Omega}{\langle \Omega \rangle}+i\pi \frac{(\delta \Omega)^{2}}{\langle \Omega \rangle^{2}} & 0\\
0 & 2-\pi^{2}\frac{(\delta \Omega)^{2}}{\langle \Omega \rangle^{2}}
\end{array} \right)
\left(\begin{array}{c} \ket{10} \\ \ket{r0}
\end{array} \right) \nonumber
\end{eqnarray}
During the $2\pi$ pulse resonant to the right atom the states $\ket{10}$ and $\ket{r0}$ change as
\begin{eqnarray}
\ket{10} \rightarrow e^{2i \frac{\Omega^{2}t}{\Delta}-i\frac{\Omega^{2}t}{\Delta}\frac{\Delta_{hf}}{\Delta}}\ket{10}, \nonumber \\
\ket{r0} \rightarrow e^{i\Delta_{vec}+2i\frac{\Omega^{2}t}{\Delta}-i\frac{\Omega^{2}t}{\Delta}\frac{\Delta_{hf}}{\Delta}}\ket{r0}, \nonumber
\end{eqnarray}
{\it i.e.} after the $2\pi$ pulse
\begin{eqnarray}
\ket{10} \rightarrow e^{4\pi i\frac{\delta \Omega}{\langle \Omega \rangle}+2\pi i\frac{(\delta \Omega)^{2}}{\langle \Omega \rangle^{2}}-i\pi \frac{\Delta_{hf}}{\Delta}}\ket{10},\nonumber \\
\ket{r0} \rightarrow e^{i\pi \frac{\Delta_{\rm vec}}{(\Omega^{2}/\Delta)}+4\pi i\frac{\delta \Omega}{\langle \Omega \rangle}+2\pi i\frac{(\delta \Omega)^{2}}{\langle \Omega \rangle^{2}}-i\pi\frac{\Delta{hf}}{\Delta}}\ket{r0}. \nonumber
\end{eqnarray}
At the end of the $2\pi$ pulse the initial state becomes
\begin{eqnarray}
\Psi \rightarrow -\frac{i}{2}e^{6\pi i\frac{\delta \Omega}{\langle \Omega \rangle}+3\pi i \frac{(\delta \Omega)^{2}}{\langle \Omega \rangle^{2}}-\frac{3\pi i\Delta_{hf}}{2\Delta}}\times \nonumber \\
\left(\begin{array}{cc}
2\pi i\frac{\delta \Omega}{\langle \Omega \rangle}+i\pi \frac{(\delta \Omega)^{2}}{\langle \Omega \rangle^{2}} & 0 \\
0 & e^{i\pi \frac{\Delta_{\rm vec}}{\Omega^{2}/\Delta}}\left(2-\pi^{2}\frac{(\delta \Omega)^{2}}{\langle \Omega \rangle^{2}}\right)
\end{array} \right)
\left(\begin{array}{c} \ket{10} \\ \ket{r0}
\end{array}\right) \nonumber
\end{eqnarray}
After the second $\pi$ pulse the states $\ket{10}$ and $\ket{r0}$ turn into
\begin{eqnarray}
\left(\begin{array}{c} \ket{10} \\ \ket{r0}
\end{array} \right)
\rightarrow -\frac{i}{2}e^{2\pi i\frac{\delta \Omega}{\langle \Omega \rangle}+i\pi \frac{(\delta \Omega)^{2}}{\langle \Omega \rangle^{2}}-\frac{i\pi \Delta_{hf}}{2\Delta}}\times \nonumber \\
\left(\begin{array}{cc}
2\pi i\frac{\delta \Omega}{\langle \Omega \rangle}+i\pi \frac{(\delta \Omega)^{2}}{\langle \Omega \rangle^{2}} & 2-\pi^{2}\frac{(\delta \Omega)^{2}}{\langle \Omega \rangle^{2}}\\
2-\pi^{2} \frac{(\delta \Omega)^{2}}{\langle \Omega \rangle^{2}} & 2\pi i\frac{\delta \Omega}{\langle \Omega \rangle}+i\pi \frac{(\delta \Omega)^{2}}{\langle \Omega \rangle^{2}}
\end{array} \right)
\left(\begin{array}{c} \ket{10} \\ \ket{r0}
\end{array} \right) \nonumber
\end{eqnarray}
If $\pi \Delta_{\rm vec}/(\Omega^{2}/\Delta)=2\pi n$, the state $\ket{10}$ at the end of the gate becomes
\begin{eqnarray}
\ket{\Psi} \rightarrow -\frac{i}{2}e^{8\pi i\frac{\delta \Omega}{\langle \Omega \rangle}+4\pi i \frac{(\delta \Omega)^{2}}{\langle \Omega \rangle^{2}}-\frac{\pi i\Delta_{hf}}{\Delta}}\times \nonumber \\
\left(\begin{array}{cc}
1-2\pi^{2} \frac{(\delta \Omega)^{2}}{\langle \Omega \rangle^{2}} & 0\\
0 & 2\pi i\frac{\delta \Omega}{\langle \Omega \rangle}+i\pi^{2}\frac{(\delta \Omega)^{2}}{\langle \Omega \rangle^{2}}
\end{array}\right)
\left(\begin{array}{c} \ket{10} \\ \ket{r0}
\end{array} \right) \nonumber
\end{eqnarray}
From $\left|-\langle 10 \vert \Psi\rangle \right|^{2}=1-4\pi^{2}(\delta \Omega)^{2}/\langle \Omega \rangle^{2}$ the error is $4\pi^{2}(\delta \Omega)^{2}/\langle \Omega \rangle^{2}$. 

Finally, the $\ket{00}$ state at the end of the gate becomes $\ket{00} \rightarrow e^{8\pi i \delta \Omega/\langle \Omega \rangle+4\pi i(\delta \Omega)^{2}/\langle \Omega \rangle^{2}-4\pi i \Delta_{hf}/\Delta}\ket{00}$, 
giving $\left|\langle 00 \vert \Psi \rangle \right|^{2}=1$. 

The error due to the variation of the Rabi frequency averaged over all initial two-qubit states is then $2\pi^{2}\left \langle (\delta \Omega)^{2}/\langle \Omega \rangle^{2}\right \rangle$, 
where the averaging in the error expression is over the ground motional atomic state.

\subsection{Excitation of atoms in minima of standing wave excitation pulse}

Finally, we analyze the undesirable excitation of atoms in "inactive" wells, situated at the minima of the standing wave excitation pulse. 

\subsubsection{$\ket{11}$ initial state}

The initial 
state $\ket{11}$ after the first $\pi$ pulse turns into
\begin{eqnarray}
\ket{\Psi} & \rightarrow & \frac{e^{i\pi \tilde{\Omega}^{2}/\Omega^{2}}}{2}\left((\ket{11}-\ket{r1})e^{-i\pi \tilde{\Omega}^{2}/2\Omega^{2}}+\right. \nonumber \\
&&\left.  +(\ket{11}+\ket{r1})e^{i\pi \tilde{\Omega}^{2}/2\Omega^{2}}\right), \nonumber 
\end{eqnarray}
where $\tilde{\Omega}$ and $\Omega$ are the Rabi frequencies at the lattice sites, corresponding to the minima and the maxima of the standing wave, and 
$\tilde{\Omega}/\Omega=\left(1-\sqrt{(1+V_{1}/4V_{0})/2}\right)/\left(1+\sqrt{(1+V_{1}/4V_{0})/2}\right)$.
After the $2\pi$ pulse resonant to the right atom, the states $\ket{11}$ and $\ket{r1}$ evolve as
\begin{eqnarray}
\ket{11} \rightarrow e^{2i\pi \tilde{\Omega}^{2}/\Omega^{2}}\times \nonumber \\
\times \left(\cos\left(\pi \tilde{\Omega}^{2}/\Omega^{2}\right)\ket{11} 
+i\sin\left(\pi \tilde{\Omega}^{2}/\Omega^{2}\right)\ket{r1}\right), \nonumber \\
\ket{r1} \rightarrow  e^{i\pi \Delta_{\rm vec}/(\Omega^{2}/\Delta)-2\pi i \tilde{\Omega}^{2}/\Omega^{2}}\ket{r1}. \nonumber
\end{eqnarray}
During the second $\pi$ pulse, resonant to the right atom, the states $\ket{11}$, $\ket{r1}$ and $\ket{1r}$ change as
\begin{eqnarray}
\left(\begin{array}{c} \ket{11} \\ \ket{r1}
\end{array} \right)
\rightarrow e^{i\pi \tilde{\Omega}^{2}/\Omega^{2}}\times  \nonumber \\
\left( \begin{array}{cc}
\cos\left(\pi \tilde{\Omega}^{2}/\Omega^{2}\right) & i\sin\left(\pi \tilde{\Omega}^{2}/\Omega^{2}\right) \\
i\sin\left(\pi \tilde{\Omega}^{2}/\Omega^{2}\right) & \cos\left(\pi \tilde{\Omega}^{2}/\Omega^{2}\right)
\end{array} \right)
\left(\begin{array}{c} \ket{11} \\ \ket{r1}
\end{array} \right) \nonumber
\end{eqnarray}
At the end of the gate the state $\ket{11}$ becomes
\begin{eqnarray}
\ket{\Psi} & = & e^{4\pi i \tilde{\Omega}^{2}/\Omega^{2}}\cos^{2}\left(\pi \tilde{\Omega}^{2}/2\Omega^{2}\right)\cos\left(\pi \tilde{\Omega}^{2}/\Omega^{2}\right)\ket{11} \nonumber \\
&& -e^{i\pi \Delta_{\rm vec}/(\Omega^{2}/\Delta)}\sin^{2}\left(\pi \tilde{\Omega}^{2}/2\Omega^{2}\right)\ket{11}+ \nonumber \\
&& +\frac{ie^{4\pi i \tilde{\Omega}^{2}/\Omega^{2}}}{4}\sin^{2}\left(2\pi \tilde{\Omega}^{2}/\Omega^{2}\right)\ket{r1}+\nonumber \\
&& +\frac{ie^{i\pi \Delta_{\rm vec}/(\Omega^{2}/\Delta)}}{2}\sin \left(\pi \tilde{\Omega}^{2}/\Omega^{2}\right)\ket{r1}+\nonumber \\
&& +ie^{2\pi i \tilde{\Omega}^{2}/\Omega^{2}-i\pi \Delta_{\rm vec}/(2\Omega^{2}/\Delta)}\times \nonumber \\
&& \times \cos \left(\pi \tilde{\Omega}^{2}/2\Omega^{2}\right)\sin \left(\pi \tilde{\Omega}^{2}/\Omega^{2}\right)\ket{1r}.  \nonumber
\end{eqnarray}
As a result, the overlap with the initial $\ket{11}$ state is
\begin{eqnarray}
\langle 11 \vert \Psi \rangle = e^{4\pi i \tilde{\Omega}^{2}/\Omega^{2}}\cos^{2}\left(\pi \tilde{\Omega}^{2}/2\Omega^{2}\right)\cos(\pi \tilde{\Omega}^{2}/\Omega^{2})\nonumber \\
-e^{i\pi \Delta_{\rm vec}/(\Omega^{2}/\Delta)}\sin^{2}\left(\pi \tilde{\Omega}^{2}/2\Omega^{2}\right). \nonumber
\end{eqnarray}
The probability that after the gate the atoms stay in $\ket{11}$ is 
\begin{eqnarray}
|\langle 11 \vert \Psi \rangle|^{2}=\cos^{4}\left(\pi \tilde{\Omega}^{2}/2\Omega^{2}\right)\cos^{2}\left(\pi \tilde{\Omega}^{2}/\Omega^{2}\right)+\sin^{4}\left(\pi \tilde{\Omega}^{2}/2\Omega^{2}\right)\nonumber \\
-\frac{1}{2}\sin^{2}\left(\pi \tilde{\Omega}^{2}/\Omega^{2}\right)\cos\left(\pi \tilde{\Omega}^{2}/\Omega^{2}\right)\cos\left(4\pi \tilde{\Omega}^{2}/\Omega^{2}\right), \nonumber 
\end{eqnarray}
where we assumed $\pi \Delta_{\rm vec}/(\Omega^{2}/\Delta)=2\pi n$. 

The ratio $\tilde{\Omega}/\Omega \approx 0.117$ for $V_{1}=V_{0}$, giving $|\langle 11 \vert \Psi \rangle|^{2}\approx 0.996$. 

\subsubsection{$\ket{01}$ initial state}

If the initial state is $\ket{01}$, it evolves into 
\begin{eqnarray}
\ket{\Psi}=e^{i\pi \tilde{\Omega}^{2}/2\Omega^{2}+i\pi \tilde{\Omega}^{2}/(2\Omega^{2})(\Delta/\Delta+\Delta_{hf})}\ket{01} \nonumber
\end{eqnarray}
after the first $\pi$ pulse resonant to the left atom.
After the $2\pi$ pulse resonant to the right atom the $\ket{01}$ state turns into
\begin{eqnarray}
\ket{01}\rightarrow e^{i\pi \tilde{\Omega}^{2}/\Omega^{2}+i\pi (\tilde{\Omega}^{2}/\Omega^{2})(\Delta/(\Delta+\Delta_{hf}))}\times \nonumber \\
\times \left(\cos\left(\pi \tilde{\Omega}^{2}/\Omega^{2}\right)\ket{01}+i\sin\left(\pi \tilde{\Omega}^{2}/\Omega^{2}\right)\ket{0r}\right). \nonumber
\end{eqnarray}
During the second $\pi$ pulse resonant to the left atom, the states $\ket{01}$ and $\ket{r0}$ change as
\begin{eqnarray}
\ket{01} \rightarrow e^{i\pi \tilde{\Omega}^{2}/2\Omega^{2}+i\pi (\tilde{\Omega}^{2}/2\Omega^{2})(\Delta/(\Delta+\Delta_{hf}))}\ket{01}, \nonumber \\
\ket{0r} \rightarrow e^{-i\pi \Delta_{\rm vec}/(2\Omega^{2}/\Delta)+i\pi \tilde{\Omega}^{2}/2\Omega^{2}+i\pi (\tilde{\Omega}^{2}/2\Omega^{2})(\Delta/(\Delta+\Delta_{hf}))}\ket{0r}. \nonumber
\end{eqnarray}
At the end of the gate
\begin{eqnarray}
\ket{\Psi}=e^{2i\pi \tilde{\Omega}^{2}/\Omega^{2}+2i\pi (\tilde{\Omega}^{2}/\Omega^{2})(\Delta/(\Delta+\Delta_{hf}))}\times \nonumber \\
\times \cos\left(\pi \tilde{\Omega}^{2}/\Omega^{2}\right)\ket{01}+\nonumber \\
+ie^{-i\pi \Delta_{\rm vec}/(\Omega^{2}/\Delta)+2i\pi \tilde{\Omega}^{2}/\Omega^{2}+2i\pi (\tilde{\Omega}^{2}/\Omega^{2})(\Delta/(\Delta+\Delta_{hf}))}\times \nonumber \\
\times \sin \left(\pi \tilde{\Omega}^{2}/\Omega^{2}\right)\ket{0r}. \nonumber
\end{eqnarray}
The overlap with the initial $\ket{01}$ state is then $\langle 01 \vert \Psi \rangle=e^{2i\pi \tilde{\Omega}^{2}/\Omega^{2}+2i\pi (\tilde{\Omega}^{2}/\Omega^{2})(\Delta/(\Delta+\Delta_{hf}))}\cos\left(\pi \tilde{\Omega}^{2}/\Omega^{2}\right)$, 
resulting in the probability to find the atomic pair in the initial state after the gate $|\langle 01 \vert \Psi\rangle|^{2}=\cos^{2}\left(\pi \tilde{\Omega}^{2}/\Omega^{2}\right)\approx 0.998$.

\subsubsection{$\ket{10}$ initial state}

The initial state $\ket{10}$ during the first $\pi$ pulse resonant to the left atom evolves into
\begin{eqnarray}
\ket{\Psi}=e^{i\pi \tilde{\Omega}^{2}/2\Omega^{2}+i\pi (\tilde{\Omega}^{2}/2\Omega^{2})(\Delta/(\Delta+\Delta_{hf}))}\times \nonumber \\
\times \left(\cos\left(\pi \tilde{\Omega}^{2}/2\Omega^{2}\right)\ket{10}+i\sin\left(\pi \tilde{\Omega}^{2}/2\Omega^{2}\right)\ket{r0}\right). \nonumber
\end{eqnarray}
During the $2\pi$ pulse, resonant to the right atom, the states $\ket{10}$ and $\ket{r0}$ change as
\begin{eqnarray}
\ket{10} \rightarrow e^{i\pi \tilde{\Omega}^{2}/\Omega^{2}+i\pi (\tilde{\Omega}^{2}/\Omega^{2})(\Delta/(\Delta+\Delta_{hf}))}\ket{10}, \nonumber \\
\ket{r0} \rightarrow e^{i\pi \Delta_{\rm vec}/(\Omega^{2}/\Delta)+i\pi \tilde{\Omega}^{2}/\Omega^{2}+i\pi (\tilde{\Omega}^{2}/\Omega^{2})(\Delta/(\Delta+\Delta_{hf}))}\ket{r0}. \nonumber
\end{eqnarray}
During the second $\pi$ pulse the states $\ket{10}$ and $\ket{r0}$ evolve as
\begin{eqnarray}
\left(\begin{array}{c} \ket{10} \\ \ket{r0}
\end{array} \right)
\rightarrow e^{i\pi \tilde{\Omega}^{2}/2\Omega^{2}+i\pi (\tilde{\Omega}^{2}/2\Omega^{2})(\Delta/\Delta+\Delta_{hf})}\times \nonumber \\
\left(\begin{array}{cc}
\cos \left(\pi \tilde{\Omega}^{2}/2\Omega^{2}\right) & i\sin \left(\pi \tilde{\Omega}^{2}/2\Omega^{2}\right) \\
i\sin \left(\pi \tilde{\Omega}^{2}/2\Omega^{2}\right) & \cos \left(\pi \tilde{\Omega}^{2}/2\Omega^{2}\right)
\end{array} \right)
\left(\begin{array}{c} \ket{10} \\ \ket{r0}
\end{array} \right) \nonumber
\end{eqnarray}
At the end of the gate the initial state $\ket{10}$ becomes
\begin{eqnarray}
\ket{\Psi}=e^{2\pi i \tilde{\Omega}^{2}/\Omega^{2}+2\pi i (\tilde{\Omega}^{2}/\Omega^{2})(\Delta/(\Delta+\Delta_{hf}))}\times \nonumber \\
\times \cos^{2}\left(\pi \tilde{\Omega}^{2}/2\Omega^{2}\right)\ket{10}\nonumber \\
-\frac{1}{4}e^{i\pi \Delta_{\rm vec}/(\Omega^{2}/\Delta)+2\pi i \tilde{\Omega}^{2}/\Omega^{2}+2\pi i (\tilde{\Omega}^{2}/\Omega^{2})(\Delta/(\Delta+\Delta_{hf}))}\times \nonumber \\
\times \sin^{2}\left(\pi \tilde{\Omega}^{2}/2\Omega^{2}\right)\ket{10}+\nonumber \\
+\frac{i}{2}e^{2\pi i \tilde{\Omega}^{2}/\Omega^{2}+2\pi i (\tilde{\Omega}^{2}/\Omega^{2})(\Delta/\Delta+\Delta_{hf})}\times \nonumber \\
\times \sin \left(\pi \tilde{\Omega}^{2}/\Omega^{2}\right)\left(1+e^{i\pi \Delta_{\rm vec}/(\Omega^{2}/\Delta)}\right)\ket{r0}, \nonumber  
\end{eqnarray}
which gives the overlap with the initial $\ket{10}$ state
\begin{eqnarray}
\langle 10 \vert \Psi \rangle =e^{2i\pi \tilde{\Omega}^{2}/2\Omega^{2}+2\pi i (\tilde{\Omega}^{2}/\Omega^{2})(\Delta/\Delta+\Delta_{hf})}\times \nonumber \\
\times \left(\cos^{2}\left(\pi \tilde{\Omega}^{2}/2\Omega^{2}\right)-e^{i\pi \Delta_{\rm vec}/(\Omega^{2}/\Delta)}\sin^{2}\left(\pi \tilde{\Omega}^{2}/2\Omega^{2}\right)\right), \nonumber
\end{eqnarray}
resulting in the probability to find the pair of atoms in the initial state $|\langle 10 \vert \Psi\rangle|^{2}=\cos^{2}\left(\pi \tilde{\Omega}^{2}/\Omega^{2}\right)$, 
where we assumed that $\pi \Delta_{\rm vec}/(\Omega^{2}/\Delta)=2\pi n$. 

The state $\ket{00}$ after the gate bcomes $\ket{\Psi}=\ket{00}e^{4\pi i (\tilde{\Omega}^{2}/\Omega^{2})(\Delta/\Delta+\Delta_{hf})}$, giving $|\langle 00 \vert \Psi \rangle|^{2}=1$. 
The probability to find the pair of atoms in the initial state after the gate, averaged over all four initial states, is $\langle P \rangle=0.998$.

\end{document}